\documentclass[a4paper, amsfonts, amssymb, amsmath, reprint, showkeys, showpacs, nofootinbib, twoside, superscriptaddress,aps,prd]{revtex4-1}
\usepackage[english]{babel}
\usepackage[utf8]{inputenc}
\usepackage[T1]{fontenc}
\usepackage{comment}
\usepackage[colorinlistoftodos, color=green!40, prependcaption]{todonotes}
\usepackage{amsthm}
\usepackage{mathtools}
\usepackage{physics}
\usepackage{xcolor}
\usepackage{graphicx}
\usepackage{adjustbox}
\usepackage{placeins}
\usepackage{csquotes}
\usepackage{enumerate}
\usepackage[caption=false]{subfig}
\usepackage{ulem}
\usepackage{hyperref}
\usepackage{orcidlink}
\usepackage{appendix}
\usepackage{silence}
\hypersetup{colorlinks=true,urlcolor=blue,citecolor=blue,linkcolor=blue,menucolor=blue,anchorcolor=blue,filecolor=blue}

\begin{document}

\title{Probing Kerr spacetime via interferometric signatures of higher-order images}

    \author{Fabiano Feleppa\,\orcidlink{0000-0002-3706-1021}}
    \email[]{ffeleppa@unisa.it}
    \affiliation{Dipartimento di Fisica ``E.R. Caianiello'', Universit\`a di Salerno, Via Giovanni Paolo II 132, I-84084 Fisciano, Italy}
    \affiliation{Istituto Nazionale di Fisica Nucleare, Sezione di Napoli, Via Cintia, 80126 Napoli, Italy}
    \affiliation{Center for Fundamental Physics, School of Physical Science and Technology, ShanghaiTech University, 393 Middle Huaxia Road, Shanghai 201210, China}

    \author{Fabio Aratore\,\orcidlink{0009-0002-0018-2513}}
    \email[]{faratore@unisa.it}
    \affiliation{Dipartimento di Fisica ``E.R. Caianiello'', Universit\`a di Salerno, Via Giovanni Paolo II 132, I-84084 Fisciano, Italy}
    \affiliation{Istituto Nazionale di Fisica Nucleare, Sezione di Napoli, Via Cintia, 80126 Napoli, Italy}

    \author{Valerio Bozza\,\orcidlink{0000-0003-4590-0136}}
    \email[]{vbozza@unisa.it}
    \affiliation{Dipartimento di Fisica ``E.R. Caianiello'', Universit\`a di Salerno, Via Giovanni Paolo II 132, I-84084 Fisciano, Italy}
    \affiliation{Istituto Nazionale di Fisica Nucleare, Sezione di Napoli, Via Cintia, 80126 Napoli, Italy}

    \author{Lorenzo Albanese\,\orcidlink{0009-0003-2259-9747}}
    \email[]{lorenzo.alba@hotmail.it}
    \affiliation{Dipartimento di Fisica ``E.R. Caianiello'', Universit\`a di Salerno, Via Giovanni Paolo II 132, I-84084 Fisciano, Italy}

\begin{abstract}
Higher-order images produced by photons that orbit close to the unstable photon region of a black hole encode information about the spacetime geometry and leave characteristic signatures accessible with long interferometric baselines. We develop an analytical description of these signatures for a compact source in the spacetime of a slowly rotating Kerr black hole. Using the strong deflection limit, we derive the complex visibility function for a source of finite size at linear order in the black-hole spin. For a static equatorial source, we obtain a direct relation between the dimensionless Kerr spin parameter and the angular displacement of successive higher-order images belonging to the same parity sequence. We then extend the construction to a compact source on a circular equatorial orbit by evaluating each image contribution at the corresponding retarded source position. Frame dragging and source motion modify the angular positions of the images and leave distinct imprints on the long-baseline visibility pattern. For the configuration considered here, these effects act in opposite angular directions, so that the resulting structure is determined by their competition. Our results provide an analytical framework for identifying the imprints of black-hole spin and source dynamics in the interferometric signatures of higher-order images.
\end{abstract}


\maketitle

\section{Introduction}
\label{sec:intro}

The Event Horizon Telescope, a very-long-baseline interferometric array operating at a wavelength of $1.3\,\mathrm{mm}$, has resolved horizon-scale emission around the supermassive black holes M87* \cite{L1,L2,L3,L4,L5,L6, Kocherlakota-2021} and Sgr~A* \cite{L12,L13,L14,L15,L16,L17}. In both cases, the observations revealed a bright, asymmetric ring surrounding a central brightness depression and allowed its characteristic angular diameter to be measured. The observed morphology and angular scale are consistent with strongly lensed emission produced in the immediate vicinity of a Kerr black hole. The publicly released Event Horizon Telescope data also enabled a direct estimate of the spin of M87* from the orbital angular momentum carried by the observed radiation \cite{Tamburini1,Tamburini2}. Related theoretical developments and applications of this observable were presented in Refs.~\cite{Tamburini3,Tamburini4,FeleppaOAM2026}.

Nevertheless, these observations resolve only the overall ring-scale morphology. General relativity predicts a much finer lensing substructure within the observed emission ring. Photons can in fact perform one or more additional orbits in the vicinity of the unstable photon region before reaching the observer, thereby generating a hierarchy of higher-order images that accumulate near the shadow boundary. For an extended emitting region, these images form a sequence of progressively narrower photon subrings, whose interferometric signatures appear at increasingly long baselines \cite{Johnson2020,GrallaLupsascaMarrone2020}. The prospect of resolving this substructure, which would provide a direct probe of the spacetime geometry, motivates the development of interferometric arrays with longer baselines and higher angular resolution. In particular, space--Earth very-long-baseline interferometry would overcome the limitation imposed by the diameter of the Earth, as envisaged by proposals such as the Black Hole Explorer \cite{LupsascaBHEX2024,JohnsonBHEX2024}, which aims to resolve the ring substructure of M87* and Sgr~A*. Measurements of photon-ring observables have also been proposed to constrain the black-hole spin \cite{Broderick-2022,Farah2026,GutierrezLara2026}.

Both numerical studies and analytical calculations of black-hole lensing are available in the literature. Here, we pursue the analytical approach, with the aim of relating the properties of higher-order images to observable interferometric signatures and thereby helping interpret future high-resolution data \cite{Gralla2020,FrostKerr2026}.

A central analytical tool for studying highly bent photon trajectories in black-hole spacetimes is the strong deflection limit. Its origin can be traced to Darwin's exact treatment of light deflection in Schwarzschild spacetime, which also yielded a logarithmic approximation for the deflection angle as the photon distance of closest approach tends to the photon-sphere radius \cite{Darwin1959}. Strongly deflected photon trajectories and their role in black-hole lensing were further studied in Refs.~\cite{Atkinson1965,Misner1973,Luminet1979,Ohanian1987,Ellis2000,Frittelli2000,Perlick2004}. This line of research has since developed into a very broad field, with extensive investigations of the strong-deflection regime, see, e.g., Refs.~\cite{Claudel2001,Hasse2002,Perlick2004-review,Iyer2007,Keeton2008,Tsupko2008, Majumdar2009,Tarasenko2010,Eiroa2011, Wei2012,Zhang2015,Alhamzawi2016,Tsukamoto2016,Aldi-Bozza-2017,Dai2018, Aratore2021,Kuang2022,Abbas2023,Duan2023,Aratore2024,Vachher2025,Tsukamoto2026,Igata20261,Igata20262,Igata20263}, including studies of higher-order
images in the form of photon rings \cite{Gralla2019,GrallaLupsasca2020,Wielgus-2021,Ayzenberg-2022,Guerrero-2022,BK-Tsupko-2022,Tsupko-2022,Eichhorn-2023, Broderick-Salehi-2023,Kocherlakota-2024-1,Kocherlakota-2024-2,Aratore-Tsupko-Perlick-2024,Tsupko2026,FrostPhotonRings2026}.

Darwin's logarithmic expansion was later organized into a systematic lensing formalism for calculating various observables in Schwarzschild spacetime \cite{Bozza2001}. The construction was then generalized to arbitrary static and spherically symmetric spacetimes \cite{Bozza2002}, and subsequently extended to Kerr black holes, first in equatorial and quasi-equatorial configurations \cite{Bozza2003,Bozza2005} and later for generic observers \cite{Bozza2006}. More recently, its range of applicability has been enlarged to include photon propagation through an inhomogeneous, cold, nonmagnetized plasma \cite{Feleppaplasma2024} and the strong deflection of massive particles in static and spherically symmetric spacetimes \cite{Feleppamassiveparticles2025}. The formalism has further been generalized to extremal rotating black holes \cite{Feleppaextremal2026}.

The properties of higher-order images obtained in the strong deflection limit can be related analytically to interferometric observables through the complex visibility function. Defined as the Fourier transform of the brightness distribution on the observer's sky, it is the primary observable of very-long-baseline interferometry. For a compact source around a spherically symmetric black hole, this connection was established in Ref.~\cite{Aratore2021}, where an analytical expression for the complex visibility was derived and the height, width, and periodicities of its staircase structure were related to the strong-deflection-limit coefficients, and hence to the properties of the underlying spacetime. This framework was subsequently applied to parametrized deviations from Schwarzschild spacetime in Ref.~\cite{Feleppa2025}. These analyses were, however, restricted to spherically symmetric geometries and assumed the orbital motion of the source to be negligible.

In this work, we address both limitations within an analytical, perturbative low-spin treatment of the Kerr spacetime. We build upon Ref.~\cite{Aratore2024}, in which the leading spin corrections to the positions and time delays of higher-order images generated by a static point source were derived. Extending this framework to a compact source of finite size, we derive an analytical expression for the complex visibility at linear order in the black-hole spin and include the circular orbital motion of the source. This allows us to determine how the visibility structure of the spherically symmetric case is modified by frame dragging and source motion, and to identify the characteristic long-baseline features produced by their combined effect.

The paper is organized as follows. In Sec.~\ref{sec:lensingKerr}, we review the Kerr lensing framework and summarize the strong-deflection-limit results for the positions and time delays of higher-order images at linear order in the black-hole spin. In Sec.~\ref{sec:visibilityfunction}, we derive the visibility function for a compact source and obtain a direct relation between the black-hole spin and the angular displacement of successive images belonging to the same parity sequence. In Sec.~\ref{sec:intsignatureimagesorbitingemitter}, we consider a compact source on a circular orbit and investigate how its dynamics and frame dragging modify the resulting visibility pattern. Finally, in Sec.~\ref{sec:conclusions}, we summarize our results and present our conclusions.

\section{Kerr lensing}
\label{sec:lensingKerr}

In this section, we outline the Kerr lensing setup, including null geodesic motion, the observer-screen map, and the critical rays defining the photon region and black-hole shadow. We then review the strong-deflection-limit results of Ref.~\cite{Aratore2024} needed for our purposes. Throughout, we use units with $G=c=1$ and signature $\{+,-,-,-\}$.

\subsection{General framework}
\label{ref:generalframework}

\begin{figure}[t!]
    \centering \includegraphics[width=0.90\columnwidth]{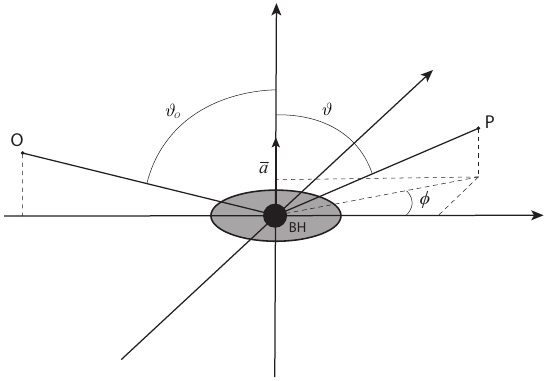}
    \caption{Schematic representation of the Kerr lensing geometry in
    Boyer--Lindquist coordinates. The black-hole spin axis is oriented along the vector $\Vec{a}$. The observer $O$ is located at polar angle $\vartheta_o$, with the origin of the azimuthal coordinate chosen such that $\phi_o=\pi$, while $P$ denotes a generic point with angular coordinates $(\vartheta,\phi)$. Figure adapted from Ref.~\cite{Bozza2006}.}
    \label{fig:Kerr_geometry}
\end{figure}

Let us consider a Kerr black hole \cite{Kerr1963} with mass $m$ and spin parameter $a$. We focus on the perturbative regime $a/m\ll 1$, in which the leading effects of the black-hole spin on the quantities of interest can be described analytically as first-order corrections in $a$. We use Boyer--Lindquist coordinates $x^\mu=(t,r,\vartheta,\phi)$, denoting the polar coordinate by $\vartheta$ in order to reserve $\theta$ for angular coordinates on the observer screen. In these coordinates, the Kerr metric reads \cite{Kerr1963,Chandrasekhar1983}
\begin{multline}
    ds^2 = \frac{\Delta - a^2\sin^2\vartheta}{\rho^2}\,dt^2 -\frac{\rho^2}{\Delta}\,dr^2 - \rho^2\,d\vartheta^2
    \\
    -\,\frac{(r^2 + a^2)^2 - a^2\Delta\sin^2\vartheta}{\rho^2}\sin^2\vartheta\,d\phi^2
    \\
    +\,\frac{4mar\sin^2\vartheta}{\rho^2}\,dt\,d\phi\,,
\label{kerr_metric}
\end{multline}
where
\begin{align}
    \Delta &= r^2 - 2mr + a^2\,,
    \\
    \rho^2 &= r^2 + a^2\cos^2\vartheta\,.
\label{kerr_functions}
\end{align}
The source is located at $(r_s,\vartheta_s,\phi_s)$, while the observer is located at $(r_o,\vartheta_o,\phi_o)$. Owing to the axial symmetry of the Kerr spacetime, only the relative azimuth between source and observer is physically relevant; we therefore choose the azimuthal origin such that $\phi_o=\pi$. The coordinate configuration adopted throughout this work is illustrated schematically in Fig.~\ref{fig:Kerr_geometry}.

Null geodesics in Kerr spacetime are characterized by two constants of motion, denoted by $J$ and $Q$:~the axial angular momentum of the photon per unit energy and the Carter constant per unit energy squared. In terms of these quantities, the geodesic equations can be written in separated form as \cite{Carter1968,Chandrasekhar1983}
\begin{gather}
    \int_\gamma \frac{dr}{\sqrt{\mathcal R(r)}} = \int_\gamma \frac{d\vartheta}{\sqrt{\Theta(\vartheta)}}\,,
\label{carter_rtheta}
    \\
    \begin{aligned}
        \phi_f - \phi_i ={}& a\int_\gamma
        \frac{r^2 + a^2 - aJ}{\Delta\sqrt{\mathcal R(r)}}\,dr - a\int_\gamma\frac{dr}{\sqrt{\mathcal R(r)}}
        \\
        &\quad+\,J\int_\gamma
        \frac{\csc^2\vartheta}{\sqrt{\Theta(\vartheta)}}\,d\vartheta\,,
        \end{aligned}
\label{carter_phi}
    \\
    \begin{aligned}
        t_f - t_i ={}& \int_\gamma\frac{r^2}{\sqrt{\mathcal R(r)}}\,dr + a^2\int_\gamma\frac{\cos^2\vartheta}{\sqrt{\Theta(\vartheta)}}\,d\vartheta
        \\
        &+\,2m\int_\gamma \frac{r(r^2 + a^2 - aJ)}{\Delta\sqrt{\mathcal R(r)}}\,dr\,,
    \end{aligned}
\label{carter_time}
\end{gather}
where
\begin{align}
    \Theta(\vartheta) &= Q + a^2\cos^2\vartheta - J^2\cot^2\vartheta\,,
\label{Theta}
    \\
    \mathcal R(r) &= \left(r^2 + a^2 - aJ\right)^2 - \Delta\left[Q + (J - a)^2\right]\,.
\label{R}
\end{align}
In Eq.~\eqref{carter_phi}, the initial and final azimuths will be identified with the source and observer azimuths, respectively. The symbol $\gamma$ denotes the photon trajectory, and all integrals in Eqs.~\eqref{carter_rtheta}--\eqref{carter_time} are evaluated along $\gamma$, with the signs of the radial and polar branches understood implicitly. Whenever the photon encounters a radial turning point, the radial integral is evaluated as the sum of the ingoing branch, from the source to the closest approach, and the outgoing branch, from the closest approach to the observer. In the same way, the angular integrals are evaluated by summing the contributions between successive polar turning points, which delimit the allowed range of the polar motion. The zeros of the functions $\mathcal R(r)$ and $\Theta(\vartheta)$ determine the radial and polar turning points, respectively.

\subsection{Critical rays, photon region, and black-hole shadow}

We now focus on the family of null geodesics that is relevant for the formation of higher-order images:~photons that pass very close to unstable photon orbits before escaping to the observer. In the Schwarzschild spacetime, because of spherical symmetry, all unstable circular photon orbits form a sphere, the photon sphere, populated by photons moving along all possible great circles. We denote the radius of the unstable photon orbits by $r_m$, which in this case takes the value
\begin{equation}
    r_m^{\rm Schw} = 3m\,.
\end{equation}
This simple picture no longer holds in Kerr spacetime. Because of frame dragging, unstable photon orbits are shifted depending on their sense of rotation:~prograde photons can orbit closer to the black hole, while retrograde photons remain farther away. Furthermore, the constant-radius photon orbits are no longer confined, in general, to a fixed plane. Rather, they provide a special case of the polar motion briefly mentioned at the end of the previous subsection:~except for the equatorial prograde and retrograde orbits, their Boyer--Lindquist radius remains constant while the polar coordinate oscillates between two turning points. Therefore, there is no unique value of $r_m$. In this case, the Schwarzschild photon sphere is replaced by a photon region, generated by a continuous family of unstable spherical photon orbits. Each member of this family is labeled by its radius $r_m$, whose value is fixed by the constants of motion of the photon \cite{Bardeen1973,Teo2003}.

As already noted, for fixed values of the constants of motion, the radial behavior of a photon is governed by the roots of $\mathcal R(r)$, which is a quartic polynomial in $r$. We define critical rays as the limiting trajectories separating photons that are captured by the black hole from photons that reach a radial turning point and escape to infinity. At this boundary, the relevant turning point becomes degenerate:~$\mathcal R(r)$ develops a double root at $r=r_m$. This is the condition for the photon to remain at constant Boyer--Lindquist radius, corresponding to one of the spherical photon orbits discussed before. Therefore
\begin{equation}
    \mathcal R(r_m) = \mathcal R'(r_m) = 0\,,
\label{critical_conditions}
\end{equation}
where the prime denotes differentiation with respect to $r$. The corresponding spherical photon orbits are unstable, as can be inferred by expanding the radial potential around the double root:~a small displacement from the critical constants of motion sends the photon away from $r_m$, either escaping to infinity or falling into the black hole. These are the photon trajectories responsible for the formation of higher-order images.

Solving Eqs.~\eqref{critical_conditions} for the constants of motion gives the critical locus parametrized by the radius $r_m$:
\begin{align}
    J_m &= \frac{r_m^3 - 3mr_m^2 + a^2r_m + ma^2}{a(m-r_m)}\,,
\label{Jm_rm}
    \\
    Q_m &= -\frac{r_m^3\left(r_m^3 - 6mr_m^2 + 9m^2r_m - 4ma^2\right)}{a^2(m - r_m)^2}\,,
\label{Qm_rm}
\end{align}
where the subscript $m$ indicates evaluation at $r_m$. The critical radius $r_m$ is restricted to the interval spanned by the unstable spherical photon orbits. The boundary values occur when the polar oscillation shrinks to zero, so that the spherical orbit reduces to an equatorial circular photon orbit. For such an orbit $\vartheta=\pi/2$ and there is no polar motion. From the differential form of the geodesic equation, $d\vartheta/d\lambda \propto \sqrt{\Theta(\vartheta)}$ \cite{Chandrasekhar1983}, with $\lambda$ an affine parameter along the geodesic, we deduce that $\Theta(\pi/2)=0$; using Eq.~\eqref{Theta}, this immediately yields $Q_m=0$. Solving this condition in the small-spin limit determines the minimum and maximum values of $r_m$ as \cite{Aratore2024}
\begin{equation}
    r_{m\pm} = 3m \pm \frac{2}{\sqrt{3}}a + \mathcal{O}\big(a^2\big)\,.
\label{rm_pm}
\end{equation}
For $a>0$, $r_{m-}$ and $r_{m+}$ correspond respectively to the prograde and retrograde equatorial photon orbits.

For a distant observer, the critical locus $(J_m,Q_m)$ can be mapped to the observer's sky. We introduce orthogonal angular coordinates $(\theta_x,\theta_y)$, centered on the black hole, with the $\theta_y$-axis aligned with the projection of the black-hole spin axis on the sky. In the asymptotic limit $r_o\gg m$, a photon observed at $(\theta_x,\theta_y)$ corresponds to screen impact parameters $(r_o\theta_x,r_o\theta_y)$. These impact parameters are directly related to the constants of motion of the photon by \cite{Bardeen1973}
\begin{align}
    r_o\theta_x &= -\frac{J}{\sqrt{1 - \mu_o^2}}\,,
\label{thetax}
    \\
    r_o\theta_y &= \pm\left[Q + \mu_o^2\left(a^2 - \frac{J^2}{1 - \mu_o^2}\right)\right]^{1/2}\,,
\label{thetay}
\end{align}
where we defined $\mu_o \coloneqq \cos\vartheta_o$. Substituting the critical locus $(J_m,Q_m)$ into Eqs.~\eqref{thetax} and \eqref{thetay} gives the shadow boundary on the observer's sky, i.e., the limiting curve approached by the higher-order images. Although this curve is naturally obtained by using $r_m$ as a parameter, it is more convenient for our purposes to label its points by their position angle on the observer screen. To this end, we first note that, to first order in $a$, the shadow boundary is the circle \cite{Aratore2024}
\begin{equation}
    \left(\theta_x - \delta\right)^2 + \theta_y^2 = \theta_{m0}^2\,,
\label{shadow_circle}
\end{equation}
with
\begin{equation}
    \theta_{m0} \coloneqq \frac{3\sqrt{3}m}{r_o}\,,
    \qquad
    \delta \coloneqq \frac{2a\sqrt{1 - \mu_o^2}}{r_o}\,.
\label{thetam0_delta_def}
\end{equation}
For $a=0$, this is the Schwarzschild circle of angular radius $\theta_{m0}$, centered on the black hole. For $a\neq0$, the circle keeps the same radius, but its center is displaced by $\delta$ along the $\theta_x$-direction; the displacement vanishes for a polar observer, $\mu_o=1$, as required by axial symmetry. For later convenience, we introduce polar coordinates $(\theta,\alpha)$ centered on the shadow rather than on the black hole:
\begin{align}
    \theta_x &= \delta + \theta\cos\alpha\,,
\label{polar_bh_x}
    \\
    \theta_y &= \theta\sin\alpha\,,
\label{polar_bh_y}
\end{align}
where $\alpha\in[0,2\pi)$ is the position angle measured
counterclockwise from the positive $\theta_x$-direction. In these coordinates, the shadow boundary is simply $\theta=\theta_{m0}$ for all $\alpha$, up to terms of order $a^2$.

Note that, in Ref.~\cite{Aratore2024}, the authors use instead polar coordinates centered on the black hole, in which the angular radius of the shadow boundary depends on the position angle. For a point close to the shadow boundary, moving the origin from the black hole to the center of the shadow changes the radial coordinate by $-\delta\cos\alpha$ and the position angle by $(\delta/\theta_{m0})\sin\alpha$, to first order in $a$. In the next subsection, we use these two relations to write the results of Ref.~\cite{Aratore2024} in the new coordinate system.

Finally, the critical radius can be expressed as a function of the position angle on the shadow boundary \cite{Aratore2024}:
\begin{equation}
    r_m = r_m(\alpha) = 3m + \frac{2}{\sqrt{3}}\,a\sqrt{1 - \mu_o^2}\cos\alpha + \mathcal{O}\big(a^2\big)\,.
\label{rm_alpha}
\end{equation}
Since the position angles measured from the black hole and from the center of the shadow differ by a term of order $a$, this expression has the same form in both cases.

\subsection{Strong deflection limit}
\label{sec:strongdeflectionlimit}

Here we give a brief summary of the strong-deflection-limit expansion of Ref.~\cite{Aratore2024}. The strong deflection limit describes the behavior of near-critical rays, whose closest approach distance $r_0$ lies close to the critical radius $r_m$ or, equivalently, whose observed direction lies close to the shadow boundary. In this regime, the radial integrals in the geodesic equations develop a logarithmic divergence as $r_0\to r_m$ \cite{Bozza2002,BozzaScarpetta2007}. This divergence is responsible for the exponential hierarchy of higher-order images, i.e., successive images accumulate exponentially close to the shadow boundary.

We parametrize the near-critical behavior directly on the observer screen. Using the polar coordinates introduced in the previous subsection, we write
\begin{align}
    \theta_x(\alpha,\epsilon) &= \delta + \theta_{m0}(1 + \epsilon)\cos\alpha\,,
\label{thetax_alpha_epsilon}
    \\
    \theta_y(\alpha,\epsilon) &= \theta_{m0}(1 + \epsilon)\sin\alpha\,,
\label{thetay_alpha_epsilon}
\end{align}
where $\epsilon\ll1$ measures the fractional angular separation from the shadow boundary.

For our purposes, the relevant lensing data are the angular positions of the higher-order images on the observer's sky and the time delays between them, both for consecutive images on the same side of the black hole and for images appearing on opposite sides. Below, we collect the main results of Ref.~\cite{Aratore2024} up to first order in the spin parameter, written in the coordinates of Eqs.~\eqref{polar_bh_x} and \eqref{polar_bh_y}; only two quantities change at this order: the first-order corrections to the position angle and to the separation from the shadow boundary.

The position angle $\alpha$ is written as
\begin{equation}
    \alpha \coloneqq \alpha_{\pm} \simeq \alpha_{0,\pm} +
    a\,\alpha_{1,\pm}\,,
\label{alpha_expansion_pm}
\end{equation}
where $\alpha_{0,\pm}$ is the zeroth-order position angle, and $a\,\alpha_{1,\pm}$ is the linear correction induced by the black-hole spin. Moreover, the subscripts $+$ and $-$ denote the two image sequences appearing on opposite sides of the black hole. The plus branch corresponds to positive-parity images (images forming on the same side as the source), while the minus branch corresponds to negative-parity ones (images forming on the opposite side from the source). In Eq.~\eqref{alpha_expansion_pm}, the angle $\alpha_{0,\pm}$ is fixed purely by the source--observer geometry; this follows from spherical symmetry in the Schwarzschild limit, where each photon trajectory lies in the plane containing the source, the black hole, and the observer. The geometrical relation is
\begin{equation}
    \cos\alpha_{0,\pm} = \frac{\mp\sqrt{1 - \mu_s^2}\sin\phi_s}{\sqrt{1 - \left(\mu_o\mu_s - \sqrt{1-\mu_o^2}\sqrt{1 - \mu_s^2}\cos\phi_s\right)^2}}\,,
\label{alpha0_pm}
\end{equation}
where we defined 
\begin{equation}
    \mu_s \coloneqq \cos\vartheta_s\,.
\end{equation}
As for $\alpha_{1,\pm}$, its two image-branch expressions are
\begin{multline}
    \alpha_{1,+} = -\tilde{\psi}_+ \frac{1 - (1 - \mu_o^2)\cos^2\alpha_{0,+}}{(-1)^{n+1}\mu_o - \mu_s\tan\alpha_{0,+}\cot\alpha_{s0,+}}
    \\
    + \frac{2\sqrt{1 - \mu_o^2}}{3\sqrt{3}\,m}\,\sin\alpha_{0,+}\,,
\label{alpha1_plus}
\end{multline}
and
\begin{multline}
    \alpha_{1,-} = -\tilde{\psi}_- \frac{1 - (1 - \mu_o^2)\cos^2\alpha_{0,-}}{(-1)^n\mu_o - \mu_s\tan\alpha_{0,-}\cot\alpha_{s0,-}}
    \\
    + \frac{2\sqrt{1 - \mu_o^2}}{3\sqrt{3}\,m}\,\sin\alpha_{0,-}\,.
\label{alpha1_minus}
\end{multline}
The first term in each of these expressions is the correction due to frame dragging obtained in Ref.~\cite{Aratore2024}. The second term is absent there:~it is the change of the position angle when the origin of the polar coordinates is moved from the black hole to the center of the shadow. Here we have introduced several quantities. The angle $\alpha_{s0,\pm}$ is defined by
\begin{equation}
    \cos\alpha_{s0,\pm} = \frac{\sqrt{1 - \mu_o^2}}{\sqrt{1 - \mu_s^2}}\cos\alpha_{0,\pm}\,,
\label{alpha_s_def}
\end{equation}
while
\begin{equation}
    \tilde{\psi}_\pm \coloneqq \frac{1}{2m}\bigg[\frac{4}{3\sqrt{3}}\psi_{0,\pm} - \log\frac{S_2(\eta_s)}{S_2(0)} - \log\frac{S_2(\eta_o)}{S_2(0)}\bigg]\,,
\label{psitilde_pm}
\end{equation}
with
\begin{align}
    \psi_{0,+} &\coloneqq (n + 1)\pi-h_+\,,
\label{psi0plus}
    \\
    \psi_{0,-} &\coloneqq n\pi + h_-\,,
\label{psi0minus}
    \\
    h_+ &\coloneqq \operatorname{arccot}\left(\frac{\sqrt{1 - \mu_s^2}\sin\alpha_{s0,+}}{\mu_s}
    \right)
    \nonumber\\
    &\quad\quad- (-1)^{n+1}\operatorname{arccot}\left(
    \frac{\sqrt{1 - \mu_o^2}\sin\alpha_{0,+}}{\mu_o}\right)\,,
\label{h_plus}
    \\
    h_- &\coloneqq - \operatorname{arccot}\left(\frac{\sqrt{1 - \mu_s^2}\sin\alpha_{s0,-}}{\mu_s}\right)
    \nonumber\\
    &\quad\quad+ (-1)^n\operatorname{arccot}\left(\frac{\sqrt{1 - \mu_o^2}\sin\alpha_{0,-}}{\mu_o}\right)\,,
\label{h_minus}
\end{align}
and
\begin{align}
    S_c(\eta_i) &\coloneqq \frac{c - \sqrt{3 - 2\eta_i}}{c + \sqrt{3 - 2\eta_i}}\,,
\label{Sc_def}
    \\
    \eta_i &\coloneqq 1 - \frac{3m}{r_i}\,, \quad i=o,s\,.
\label{etaidef}
\end{align}
Note that, in the strong-deflection formalism, the finite-distance variable $\eta_i$ is commonly defined in terms of the spin-dependent Kerr critical radius $r_m$ of Eq.~\eqref{rm_alpha}. Here, as in Ref.~\cite{Aratore2024}, it is evaluated using the Schwarzschild critical radius, $r_m^{\rm Schw}=3m$. At first order in $a$, this only neglects, in the separation of the images from the shadow boundary, a correction smaller than the one retained below by a factor of order $m/r_i$. Moreover, although we retain $\eta_o$ explicitly in the formulas, in all applications below we take $\eta_o\simeq1$, corresponding to an asymptotically distant observer.

In the above formulas, $n$ counts the number of half-orbits performed by photons around the black hole. Consecutive values of $n$ therefore correspond to images appearing on opposite sides, while images on the same side are related by $n\to n+2$. In our half-orbit labeling, the first higher-order image is generated by a photon trajectory that completes one full loop around the black hole, and therefore has $n=2$. Following the ordering of the two branches in Ref.~\cite{Aratore2024}, the negative-parity image in each opposite-side pair performs one additional half-orbit with respect to the corresponding positive-parity image. Therefore, positive-parity images carry even values of $n$, while negative-parity images carry odd values of $n$. Notice that in Ref.~\cite{Aratore2024} the authors adopt a different convention, in which the integer $n$ denotes the number of inversions in the polar motion. Therefore, the integer $n$ used in Ref.~\cite{Aratore2024} must be replaced here by $n+1$ for the positive-parity branch and by $n$ for the negative-parity branch. This correspondence, and the expressions above, hold for $\mu_o>0$ and a source behind the black hole, as in all the configurations considered below, with $\alpha_{0,-}=\alpha_{0,+}+\pi$ and $\alpha_{s0,-}=\alpha_{s0,+}+\pi$.

The relative angular separation from the Kerr shadow boundary is written as
\begin{equation}
    \epsilon_{\pm} \simeq \epsilon_{0,\pm} + a\,\epsilon_{1,\pm}\,,
\end{equation}
where
\begin{align}
    \epsilon_{0,\pm} &= 216\,S_{\sqrt{3}}(\eta_s)S_{\sqrt{3}}(\eta_o)e^{-\psi_{0,\pm}}\,,
\label{epsilon0pm}
    \\
   \epsilon_{1,\pm} &= -\epsilon_{0,\pm}\sqrt{1-\mu_o^2}\,\cos\alpha_{0,\pm}\,\tilde{\psi}_\pm\,.
\label{epsilon1pm}
\end{align}
Note that, in Ref.~\cite{Aratore2024}, $\epsilon_{1,\pm}$ contains an additional term proportional to $1/(\sqrt{3}m)$, which drops out when the separation is measured in units of the constant radius $\theta_{m0}$, as done here.

The angular coordinates of higher-order images on the observer's sky are then given by
\begin{align}
    \theta_{x,\pm} &= \delta + \theta_{m0}\left(1 + \epsilon_\pm\right)\cos\alpha_\pm\,,
\label{image_theta1_pm}
    \\
    \theta_{y,\pm} &= \theta_{m0}\left(1 + \epsilon_\pm\right)
    \sin\alpha_\pm\,.
\label{image_theta2_pm}
\end{align}
Accordingly, the radial angular position of the image is
\begin{equation}
    \theta_\pm = \theta_{m0}\left(1+\epsilon_\pm\right) \simeq \theta_{0,\pm} + a\,\theta_{1,\pm}\,,
\label{theta_radial_pm}
\end{equation}
where
\begin{align}
    \theta_{0,\pm} &= \theta_{m0}\left(1 + \epsilon_{0,\pm}\right)\,,
\label{theta0pm}
    \\
    \theta_{1,\pm} &= \theta_{m0}\,\epsilon_{1,\pm}\,.
\label{theta1pm}
\end{align}

Finally, the time delay between two consecutive images on the same side of the black hole is
\begin{equation}
    \Delta t_\pm^{\rm ss} \simeq \Delta t_0^{\rm ss} + a\,\Delta t_{1,\pm}^{\rm ss}\,,
\label{dt_same_n}
\end{equation}
where
\begin{align}
    \Delta t_0^{\rm ss} &= 6\sqrt{3}\pi m\,,
    \\
    \Delta t_{1,\pm}^{\rm ss} &= 4\pi \sqrt{1 - \mu_o^2}\cos\alpha_{0,\pm}\,,
\end{align}
while the delay between opposite-side images, written using the positive-parity image as reference, is
\begin{equation}
    \Delta t^{\rm os} \simeq \Delta t^{\rm os}_0 + a\,\Delta t^{\rm os}_1\,,
\label{dt_opp_n}
\end{equation}
with
\begin{align}
    \Delta t^{\rm os}_0 &= 6\sqrt{3}mh_+\,,
    \\
    \Delta t^{\rm os}_1 &= -\sqrt{1 - \mu_o^2}\cos\alpha_{0,+}\bigg[4\pi(n + 1)
    \nonumber\\
    &\quad\quad -\,3\sqrt{3}\left(\log\frac{S_2(\eta_s)}{S_2(0)} + \log\frac{S_2(\eta_o)}{S_2(0)}\right)\bigg]\,.
\end{align}
In these expressions, ``ss'' stands for ``same side'', while ``os'' stands for ``opposite side''. The logarithmic terms in $\Delta t^{\rm os}_1$, which are absent in Eq.~(6.7) of Ref.~\cite{Aratore2024}, follow from Eqs.~(6.2) and (6.5) therein. They cancel in the same-side delay, but not in the opposite-side one, since $\cos\alpha_{0,-}=-\cos\alpha_{0,+}$.

This concludes our summary of the Kerr lensing ingredients needed in the next section.

\section{Complex visibility function}
\label{sec:visibilityfunction}

In radio interferometry, the complex visibility function provides the Fourier-space description of the image observed on the sky. For a given source with surface brightness $I(\theta_x,\theta_y)$, written in terms of the angular screen coordinates introduced in the previous section, the visibility is given by \cite{Thompson2017,Aratore2021}
\begin{equation}
    V(u,v) = \int d\theta_x\,d\theta_y\,
    I(\theta_x,\theta_y)e^{-2\pi i(u\theta_x + v\theta_y)}\,.
\label{visibility_def_cartesian}
\end{equation}
Here $(u,v)$ are the spatial-frequency coordinates sampled by an
interferometric baseline, measured in units of the observing wavelength. The modulus $|V(u,v)|$ gives the visibility amplitude at the sampled spatial frequency, while the phase of $V(u,v)$ is sensitive to the position and asymmetry of the emission on the observer's sky. In this Fourier-space representation, extended image components contribute to a large extent at low spatial frequencies, whereas compact image features produce appreciable visibility amplitudes at high spatial frequencies. Since high spatial frequencies are sampled by long projected baselines, the higher-order images considered in this work leave their imprint in the long-baseline behavior of the visibility function \cite{Aratore2021,Johnson2020,Jia2024}.

In the polar coordinates of Eqs.~\eqref{polar_bh_x} and \eqref{polar_bh_y}, Eq.~\eqref{visibility_def_cartesian} becomes
\begin{multline}
    V(u,v) = e^{-2\pi iu\delta}\int_0^{2\pi}d\alpha
    \int_0^\infty d\theta\,\theta\,
    I(\theta,\alpha)
    \\
    \times e^{-2\pi i\theta(u\cos\alpha + v\sin\alpha)}\,.
\label{visibility_def_polar}
\end{multline}
The phase factor $e^{-2\pi iu\delta}$, which comes from the displacement of the origin of the polar coordinates from the black hole, is common to all images and does not affect the visibility amplitude.

We consider a source with finite but small extension around the central position $(r_s,\vartheta_s,\phi_s)$. Different points of the source are mapped by the lensing equations summarized in the previous section to nearby points on the observer's sky, so each higher-order image acquires a finite size. As we shall see below, the assumption that the source is compact allows us to describe each image locally by its central position and by the first-order variation of this position across the source. Each image has radial angular thickness $\Delta\theta_\pm$ and angular extension $\Delta\alpha_\pm$ along the arc, i.e., tangential angular size $\theta_\pm|\Delta\alpha_\pm|$.

For definiteness, we take the local brightness profile of each higher-order image to be Gaussian in the radial and angular directions on the observer screen. This factorized form is possible because, in the coordinates of Eqs.~\eqref{polar_bh_x} and \eqref{polar_bh_y}, each image is a thin arc of the circle $\theta=\theta_\pm$. In polar coordinates centered on the black hole, instead, the radial position of the arc would change along its length by an amount of first order in $a$, which is not small compared with its radial thickness. The contribution of all higher-order images to the surface brightness is then written as
\begin{multline}
    I(\theta,\alpha) = I_0\sum_{\pm}
    \sum_{n\in\mathcal{N}_\pm}
    \exp\left[-\frac{(\theta - \theta_\pm)^2}{2\Delta\theta_\pm^2}\right]
    \\
    \times \exp\left[-\frac{(\alpha - \alpha_\pm)^2}{2\Delta\alpha_\pm^2}\right]\,,
\label{image_intensity_gaussian}
\end{multline}
where $I_0$ is the central surface brightness of the source, which is conserved by gravitational lensing. The half-orbit label $n$ runs over
\begin{equation}
    \mathcal{N}_+ = \{2,4,6,\ldots\}\,,
    \qquad
    \mathcal{N}_- = \{3,5,7,\ldots\}\,.
\label{Npm_def}
\end{equation}
The $+$ and $-$ branches contain the positive- and negative-parity images, respectively, and in each term $\theta_\pm$, $\alpha_\pm$, $\Delta\theta_\pm$, and $\Delta\alpha_\pm$ are evaluated at the corresponding $n$.

Let $(\Delta r_s,\Delta\vartheta_s,\Delta\phi_s)$ denote the characteristic extensions of the source around its central position. To linear order in the source size, the corresponding angular extension of each higher-order image is obtained from the total variation of $\alpha_\pm$:
\begin{equation}
    \Delta\alpha_\pm \simeq \frac{\partial\alpha_\pm}{\partial r_s}\Delta r_s + \frac{\partial\alpha_\pm}{\partial\vartheta_s}\Delta\vartheta_s + \frac{\partial\alpha_\pm}{\partial\phi_s}\Delta\phi_s\,.
\label{Delta_alpha_pm_def}
\end{equation}
Although the above expression is written in terms of the physical source-size parameters, the lensing quantities presented in Sec.~\ref{sec:strongdeflectionlimit} depend on the source coordinates through the variables
\begin{equation}
    \eta_s = 1 - \frac{3m}{r_s}\,,
    \qquad
    \mu_s = \cos\vartheta_s\,,
    \qquad
    \phi_s\,.
\end{equation}
It is therefore convenient to express the first-order variations in this basis, writing
\begin{equation}
    \Delta\eta_s = \frac{3m}{r_s^2}\,\Delta r_s\,,
    \qquad
    \Delta\mu_s = -\sin\vartheta_s\,\Delta\vartheta_s\,,
\label{source_size_change_variables}
\end{equation}
while $\Delta\phi_s$ is unchanged. The total variation in Eq.~\eqref{Delta_alpha_pm_def} can then equivalently be written as
\begin{equation}
    \Delta\alpha_\pm = \frac{\partial\alpha_\pm}{\partial\eta_s}\Delta\eta_s + \frac{\partial\alpha_\pm}{\partial\mu_s}\Delta\mu_s + \frac{\partial\alpha_\pm}{\partial\phi_s}\Delta\phi_s\,.
\label{Delta_alpha_pm_def2}
\end{equation}
Using Eq.~\eqref{alpha_expansion_pm}, the above equation reads
\begin{equation}
    \Delta\alpha_\pm = \Delta\alpha_{0,\pm} + a\,\Delta\alpha_{1,\pm}\,,
\label{Delta_alpha_pm_expansion}
\end{equation}
where
\begin{align}
    \Delta \alpha_{0,\pm} &= \frac{\partial\alpha_{0,\pm}}{\partial\mu_s}\Delta\mu_s + \frac{\partial\alpha_{0,\pm}}{\partial\phi_s}\Delta\phi_s\,,
\label{Delta_alpha0pm}
    \\
    \Delta \alpha_{1,\pm} &= \frac{\partial\alpha_{1,\pm}}{\partial\eta_s}\Delta\eta_s + \frac{\partial\alpha_{1,\pm}}{\partial\mu_s}\Delta\mu_s + \frac{\partial\alpha_{1,\pm}}{\partial\phi_s}\Delta\phi_s\,.
\label{Delta_alpha1pm}
\end{align}

Similarly, using Eqs.~\eqref{theta_radial_pm}, \eqref{theta0pm}, and \eqref{theta1pm}, the radial thickness of the image is written as
\begin{equation}
    \Delta\theta_\pm \simeq \theta_{m0}\,\Delta\epsilon_{0,\pm} + a\,\theta_{m0}\,\Delta\epsilon_{1,\pm}\,,
\label{Delta_theta_pm_def}
\end{equation}
where
\begin{equation}
    \Delta\epsilon_{i,\pm} \simeq
    \frac{\partial\epsilon_{i,\pm}}{\partial\eta_s}\Delta\eta_s + \frac{\partial\epsilon_{i,\pm}}{\partial\mu_s}\Delta\mu_s + \frac{\partial\epsilon_{i,\pm}}{\partial\phi_s}\Delta\phi_s\,,
\label{Delta_epsilon_pm_def}
\end{equation}
for $i=0,1$.

Our goal is now to obtain an analytical expression for the visibility function. To this end, we start by substituting Eq.~\eqref{image_intensity_gaussian} into Eq.~\eqref{visibility_def_polar}, and perform the radial integral first. For each higher-order image, the radial profile is centered at
$\theta=\theta_\pm$ and has width $\Delta\theta_\pm$. In the strong deflection limit considered here, the image centers approach the shadow boundary exponentially with $n$; indeed, $\theta_\pm=\theta_{m0}(1+\epsilon_\pm)$, with $\epsilon_\pm\ll1$. The radial width produced by a finite source is obtained by varying this relation across the source and is therefore controlled by the corresponding variation of $\epsilon_\pm$. Since this variation is small compared with the angular scale set
by $\theta_\pm$, the higher-order images are narrow in the radial direction, $\Delta\theta_\pm\ll\theta_\pm$. The Gaussian factor therefore localizes the radial integral to a small neighborhood of $\theta_\pm$. We can then extend the integration range from $[0,\infty)$ to $(-\infty,\infty)$, adding only exponentially small Gaussian tails, which are negligible at the order considered here. This replacement is useful because it reduces the radial integration to the standard Fourier transform of a Gaussian, obtaining
\begin{align}
    &\int_{-\infty}^\infty d\theta\,\theta\,\exp\left[
    -\frac{(\theta-\theta_\pm)^2}{2\Delta\theta_\pm^2}\right]
    e^{-2\pi i\theta(u\cos\alpha + v\sin\alpha)}
    \nonumber\\
    &\qquad\qquad\simeq \sqrt{2\pi}\,
    \theta_\pm|\Delta\theta_\pm|\,
    e^{-2\pi i\theta_\pm(u\cos\alpha + v\sin\alpha)}
    \nonumber\\
    &\qquad\qquad\qquad \times\,e^{-2\pi^2\Delta\theta_\pm^2
    (u\cos\alpha + v\sin\alpha)^2}\,,
\label{theta_integral_result}
\end{align}
where we evaluated the Jacobian factor $\theta$ at $\theta_\pm$. Hence the visibility reduces to
\begin{multline}
    V(u,v) \simeq \sqrt{2\pi}\,I_0\,e^{-2\pi iu\delta}
    \sum_{\pm}\sum_{n\in\mathcal{N}_\pm}
    \theta_\pm|\Delta\theta_\pm|
    \\
    \times \int_0^{2\pi} d\alpha\exp\left[-\frac{(\alpha-\alpha_\pm)^2}{2\Delta\alpha_\pm^2}\right]
    e^{-2\pi i\theta_\pm(u\cos\alpha + v\sin\alpha)}
    \\
    \times e^{-2\pi^2\Delta\theta_\pm^2
    (u\cos\alpha + v\sin\alpha)^2}\,.
\label{visibility_after_theta_integral}
\end{multline}
After the radial integration, we are left with an integral over the angular coordinate $\alpha$. Since the source is assumed to be compact, the corresponding image occupies only a small angular interval around its central position $\alpha_\pm$. The integral is therefore dominated by values of $\alpha$ close to $\alpha_\pm$. The Gaussian angular
profile suppresses the integrand away from this neighborhood, so the angular integration range can also be extended to $(-\infty,\infty)$, analogously to the radial integral. We then linearize the combination $u\cos\alpha+v\sin\alpha$ around $\alpha=\alpha_\pm$. Together with the quadratic angular Gaussian, this approximation makes the complete exponent quadratic in $\alpha-\alpha_\pm$, so that the integral can be evaluated analytically, yielding
\begin{multline}
    V(u,v) \simeq 2\pi I_0\,e^{-2\pi iu\delta}\sum_{\pm}
    \sum_{n\in\mathcal{N}_\pm}\theta_\pm|\Delta\theta_\pm\Delta\alpha_\pm|
    \\
    \times e^{-\pi^2\mathcal{R}_\pm(u,v)}\,e^{-2\pi i\theta_\pm\mathcal{J}_\pm(u,v)}\,.
\label{visibility_final}
\end{multline}
The function $\mathcal{R}_\pm$ is given by
\begin{align}
    \mathcal{R}_\pm(u,v) &= (u^2 + v^2)
    \left(\Delta\theta_\pm^2 + \theta_\pm^2\Delta\alpha_\pm^2\right)
    \nonumber\\
    &\quad+ (u^2 - v^2)\left(\Delta\theta_\pm^2 - \theta_\pm^2\Delta\alpha_\pm^2\right)\cos 2\alpha_\pm
    \nonumber\\
    &\quad+ 2uv\left(\Delta\theta_\pm^2 - \theta_\pm^2\Delta\alpha_\pm^2\right)\sin 2\alpha_\pm\,,
\label{R_pm_def}
\end{align}
while the function $\mathcal{J}_\pm$ is
\begin{multline}
    \mathcal{J}_\pm(u,v) = \left[1 - \pi^2(u^2 + v^2)\Delta\theta_\pm^2\Delta\alpha_\pm^2
    \right]
    \\
    \hspace{-2cm}\times \left(u\cos\alpha_\pm + v\sin\alpha_\pm\right) + \pi^2\Delta\theta_\pm^2\Delta\alpha_\pm^2
    \\
    \times \big[u(u^2 - 3v^2)\cos 3\alpha_\pm + v(3u^2 - v^2)\sin 3\alpha_\pm\big]\,.
\label{J_pm_def}
\end{multline}
Equation~\eqref{visibility_final}, together with the definitions of $\mathcal{R}_\pm$ and $\mathcal{J}_\pm$ in Eqs.~\eqref{R_pm_def} and \eqref{J_pm_def}, provides the complex visibility function for the higher-order images of a compact source in the Kerr spacetime considered here. In Eq.~\eqref{visibility_final}, we neglect corrections proportional to $\Delta\theta_\pm^2\Delta\alpha_\pm^2$ in the amplitude but retain them in the phase, where their effect can be enhanced at long baselines. Note that a sufficient order-of-magnitude condition for neglecting arc curvature when linearizing $u\cos\alpha+v\sin\alpha$ is $\pi\theta_\pm B\,\Delta\alpha_\pm^2 \ll 1$, where $B=\sqrt{u^2+v^2}$ is the baseline length.

\subsection{Interlude:~a spin formula from same-side image displacements}
\label{sec:spinformula}

We now derive a direct expression for the dimensionless spin parameter $a/m$ in terms of the angular displacement between two successive higher-order images belonging to the same parity sequence. The derivation applies to a static equatorial source.

In the formulas presented in the previous sections, the dependence on $n$ has generally been left implicit. However, for the purpose of the present derivation, we make it explicit by writing $\alpha_\pm(n)$, $\tilde{\psi}_\pm(n)$, and $\psi_{0,\pm}(n)$.

Consider two successive images belonging to the same parity sequence, with half-orbit numbers $n$ and $n+2$. For a static source, the zeroth-order image angle $\alpha_{0,\pm}$ does not depend on $n$. Moreover, for an equatorial source, $\mu_s=0$, and Eqs.~\eqref{alpha1_plus} and \eqref{alpha1_minus} give
\begin{multline}
    \alpha_\pm(n) \simeq \alpha_{0,\pm} + a\bigg[\frac{1 - (1 - \mu_o^2)\cos^2\alpha_{0,\pm}}{\mu_o}\tilde{\psi}_\pm(n)
    \\
    + \frac{2\sqrt{1 - \mu_o^2}}{3\sqrt{3}\,m}\,\sin\alpha_{0,\pm}\bigg]\,.
\end{multline}
Here we have used the fact that $n$ is even for the positive-parity sequence and odd for the negative-parity sequence.

Under $n\rightarrow n+2$, the parity of the half-orbit number does not change. Therefore, the factors $(-1)^n$ and $(-1)^{n+1}$ entering $h_-$ and $h_+$, respectively, remain unchanged. Since, for a static source, the remaining quantities entering $h_\pm$ depend only on the source--observer geometry, we have
\begin{equation}
    h_\pm(n+2) = h_\pm(n)\,.
\end{equation}
\begin{figure}[t!]
    \centering \includegraphics[width=0.98\columnwidth]{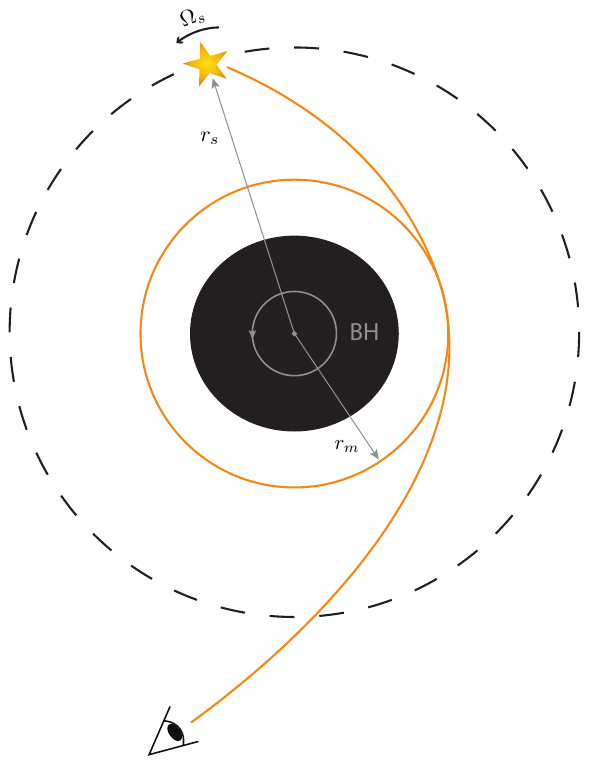}
    \caption{Polar view of the orbiting-source configuration. The dashed circle represents the circular equatorial orbit of radial coordinate $r_s$, along which the compact source moves with angular velocity $\Omega_s$ around the Kerr black hole. The orange curve represents a possible photon trajectory passing close to the unstable photon orbit at the critical radius $r_m$.}
\label{fig:orbiting_source_geometry}
\end{figure}
Likewise, the logarithmic finite-distance contributions
\begin{equation}
    \log\frac{S_2(\eta_s)}{S_2(0)}
    \quad \text{and} \quad
    \log\frac{S_2(\eta_o)}{S_2(0)}
\end{equation}
entering $\tilde{\psi}_\pm(n)$ depend only on the radial positions of the source and observer, through $\eta_s$ and $\eta_o$, and are therefore independent of $n$. The entire dependence of $\tilde{\psi}_\pm(n)$ on the image order is consequently contained in $\psi_{0,\pm}(n)$. Using Eqs.~\eqref{psi0plus} and \eqref{psi0minus}, for either parity sequence, one obtains
\begin{equation}
    \psi_{0,\pm}(n+2) - \psi_{0,\pm}(n) = 2\pi\,.
\end{equation}
It therefore follows from Eq.~\eqref{psitilde_pm} that
\begin{equation}
    \tilde{\psi}_\pm(n+2) - \tilde{\psi}_\pm(n) = \frac{4\pi}{3\sqrt{3}\,m}\,.
\end{equation}
Consequently, we can write
\begin{equation}
    \alpha_\pm(n+2) - \alpha_\pm(n) \simeq \frac{4\pi}{3\sqrt{3}}\,\frac{1 - (1 - \mu_o^2)\cos^2\alpha_{0,\pm}}{\mu_o}\frac{a}{m}\,.
\label{diffalpha_same_side}
\end{equation}
To linear order, the dimensionless Kerr spin parameter is then given by
\begin{equation}
    \frac{a}{m} = \frac{3\sqrt{3}}{4\pi}\,\frac{\mu_o}{1 - (1 - \mu_o^2)\cos^2\alpha_{0,\pm}}\,\left[\alpha_\pm(n+2) - \alpha_\pm(n)\right]\,.
\label{spinformula}
\end{equation}
Note that the last term of $\alpha_\pm(n)$ does not depend on $n$ and cancels in the difference.

The above derivation holds for a static source. In this case, all images belonging to a given parity sequence correspond to the same source position, so that the corresponding zeroth-order image angles $\alpha_{0,\pm}$ are identical and cancel in the difference between the images with labels $n$ and $n+2$.

For a moving source, instead, the source position entering each image order must be evaluated at the corresponding retarded emission time, as we will discuss in the next section. The zeroth-order image angles associated with $n$ and $n+2$ therefore generally refer to different source positions and no longer cancel directly. Extending the above derivation to moving emitters would require knowledge of the orbital dynamics of the source and of the relative propagation delay between the two image orders. The images would then have to be compared at arrival times separated by this delay. Since the required times will generally not coincide with those at which the interferometric data are sampled, an interpolation of the time-dependent visibility function would generally be needed. Such an analysis is beyond the scope of the present work.

\section{Interferometric signature of an orbiting compact source}
\label{sec:intsignatureimagesorbitingemitter}

We now consider the case in which the compact emitter moves on an orbit around the black hole. The complex visibility formula derived in the previous section depends on the source position through the lensing quantities $\theta_\pm$, $\alpha_\pm$, $\Delta\theta_\pm$, and $\Delta\alpha_\pm$. When the emitter is in motion, the source position entering the contribution of a given higher-order image must be evaluated at the emission time of the photons forming that image, which differs from one image to another. Therefore, in each term of the visibility function, the source azimuth entering the lensing quantities must be replaced by the corresponding retarded source azimuth, i.e., the azimuth of the source at the earlier time when these photons were emitted. We make this retarded-source prescription explicit below.

For simplicity, we consider a compact source on a circular equatorial orbit of radius $r_s$. Its polar coordinate is then fixed to
\begin{equation}
    \vartheta_s = \frac{\pi}{2}\,.
\end{equation}
A polar view of the physical configuration considered here is shown
schematically in Fig.~\ref{fig:orbiting_source_geometry}.

The angular velocity of the source is \cite{Bardeen1972}
\begin{equation}
    \Omega_s = \pm \frac{m^{1/2}}{r_s^{3/2} \pm am^{1/2}}\,,
\end{equation}
with the upper sign referring to prograde orbits and the lower sign to retrograde orbits. Therefore, the source azimuthal position evolves as
\begin{equation}
    \phi_s(t) = \phi_{s,0} + \Omega_s t\,,
\label{source_orbit_phi}
\end{equation}
where $t$ is the coordinate time and $\phi_{s,0}$ is the source azimuth at $t=0$. For a prograde circular orbit in a slowly rotating Kerr spacetime, the angular velocity can be expanded in the spin parameter as
\begin{equation}
    \Omega_s \simeq \Omega_{s0} + a\,\Omega_{s1}\,,
\label{Omega_s_expansion}
\end{equation}
with
\begin{equation}
    \Omega_{s0} = \sqrt{\frac{m}{r_s^3}}\,,
    \qquad
    \Omega_{s1} = -\frac{m}{r_s^3}\,.
\label{Omega_s_coefficients}
\end{equation}
The first term is the Schwarzschild Keplerian angular velocity, while the linear term gives the leading correction due to the black-hole spin.

As previously anticipated, for an orbiting emitter, the lensing quantities associated with each higher-order image must be evaluated at the source azimuth at the emission time of the photons forming that image. We denote the retarded source azimuth associated with the image labeled by $n$ as $\phi_{s,n}(t)$. Since even and odd values of $n$ belong to $\mathcal{N}_+$ and $\mathcal{N}_-$, respectively, the image branch is already specified by $n$ and no additional $\pm$ label is required. Since the source moves with angular velocity $\Omega_s$, a propagation-time difference $\Delta t$ corresponds to an azimuthal displacement $\Omega_s\Delta t$. Therefore, if a photon was emitted a time $\Delta t$ earlier, the source azimuth entering the lensing formulas is shifted backward by $\Omega_s\Delta t$. Since the strong deflection limit is accurate starting from the first higher-order image, $n=2$, we take this image as the reference for the retarded-azimuth construction. Accordingly, we label each observed snapshot by the emission time $t$ of its reference image $n=2$, so that $\phi_{s,2}(t)=\phi_s(t)$, with $\phi_s(t)$ given by Eq.~\eqref{source_orbit_phi}.

The orbiting-source visibility is therefore
\begin{multline}
    V(u,v;t) \simeq 2\pi I_0\,e^{-2\pi iu\delta}\sum_{\pm}
    \sum_{n\in\mathcal{N}_\pm}\bigg[ \theta_\pm|\Delta\theta_\pm\Delta\alpha_\pm|
    \\
    \times e^{-\pi^2\mathcal{R}_\pm(u,v)}\,e^{-2\pi i\theta_\pm\mathcal{J}_\pm(u,v)}
    \bigg]_{\phi_s=\phi_{s,n}(t)}\,.
\label{visibility_orbiting_source}
\end{multline}
All quantities inside the square brackets of the above expression are the same lensing quantities defined in the previous section. The only change is that, for each image, the source azimuth entering $\theta_\pm$, $\alpha_\pm$, $\Delta\theta_\pm$, and $\Delta\alpha_\pm$ is the retarded azimuth $\phi_{s,n}(t)$. We also keep the same $I_0$ for all images. For a moving source, the observed surface brightness of each image also contains the Doppler factor at emission, which differs between images because their photons leave the source in different directions relative to its velocity. For the orbit considered in Sec.~\ref{sec:application}, this changes the relative brightness of the images by a few percent and has no visible effect on the visibility maps.

\subsection{Retarded azimuths in the static-delay approximation}
\label{sec:static_delay}

It remains to determine the retarded azimuths $\phi_{s,n}(t)$. To this end, we use the delays of Sec.~\ref{sec:strongdeflectionlimit}, which depend on the source position through the quantities $h_+$ and $\alpha_{0,\pm}$, and evaluate all of them at the same azimuth, namely the reference azimuth $\phi_s(t)$. This is an approximation, since the propagation time of the photons forming each image observed at time $t$ should be evaluated at the azimuth from which its photons were emitted, i.e., at its own retarded azimuth $\phi_{s,n}(t)$. It amounts to computing all the delays as if the source were at rest at $\phi_s(t)$, and we therefore refer to it as the static-delay approximation. Its advantage is that it yields closed-form expressions for $\phi_{s,n}(t)$. The exact retarded azimuths are derived in Sec.~\ref{sec:exact_delay}, and the accuracy of this approximation is discussed in Sec.~\ref{sec:application}.

Let us now consider the retarded source azimuth for the positive-parity sequence. Since two half-orbits correspond to one full loop, the positive-parity image with label $n$ contains $(n-2)/2$ more full-loop intervals than the $n=2$ image. Since each full-loop interval on the positive-parity sequence contributes the same-side delay $\Delta t^{\rm ss}_+$, we can write
\begin{equation}
    \phi_{s,n}(t) = \phi_s(t) - \frac{n - 2}{2}\,\Omega_s\,\Delta t^{\rm ss}_+\,,
    \quad
    n \in \mathcal{N}_+\,.
\label{phi_ret_plus_exact}
\end{equation}

We now consider the retarded source azimuth $\phi_{s,n}(t)$ for $n\in\mathcal{N}_-$. In this case, the $n$-labeling requires some care because the opposite-side delay in Eq.~\eqref{dt_opp_n} is written using the positive-parity image as reference. Thus, the integer $n$ appearing in Eq.~\eqref{dt_opp_n} labels the positive-parity member of an opposite-side pair. A negative-parity image with label $n$ is instead paired with the positive-parity image with label $n-1$. As a consequence, when constructing $\phi_{s,n}(t)$, for $n\in\mathcal{N}_-$, the opposite-side delay must be evaluated with the replacement $n\to n-1$. Notice that this replacement applies to the entire expression for $\Delta t^{\rm os}$, including the dependence contained in the quantity $h_+$.

Therefore, in order to obtain the retarded source azimuth of the negative-parity image, we first take the retarded source azimuth of the paired positive-parity image, $\phi_{s,n-1}(t)$, and then subtract the additional opposite-side delay:
\begin{equation}
    \phi_{s,n}(t) = \phi_{s,n-1}(t) - \Omega_s\left.\Delta t^{\rm os}\right|_{n\to n-1}\,,
    \quad
    n\in\mathcal{N}_-\,.
\label{phi_ret_minus_from_plus}
\end{equation}
Equivalently, substituting Eq.~\eqref{phi_ret_plus_exact} with
\(n\to n-1\), one obtains
\begin{multline}
    \phi_{s,n}(t) = \phi_s(t) -
    \Omega_s\left[\frac{n - 3}{2}\Delta t_+^{\rm ss}\right.
    \\
    \left.+ \left.\Delta t^{\rm os}
    \right|_{n\to n-1}\right]\,, \quad n\in \mathcal{N}_-\,.
\label{phi_ret_minus}
\end{multline}
Now, using the first-order expansions in Eqs.~\eqref{dt_same_n}, \eqref{dt_opp_n}, and \eqref{Omega_s_expansion}, together with Eq.~\eqref{source_orbit_phi}, Eqs.~\eqref{phi_ret_plus_exact} and
\eqref{phi_ret_minus} become
\begin{align}
    \phi_{s,n}(t) &\simeq \phi_{s,0} + \Omega_{s0} t - \frac{n - 2}{2}\Omega_{s0}\Delta t_0^{\rm ss}
    \nonumber\\
    &+\,a\left[\Omega_{s1} t - \frac{n - 2}{2}\left(
    \Omega_{s0}\Delta t_{1,+}^{\rm ss} + \Omega_{s1}\Delta t_0^{\rm ss}\right)\right]\,,
\label{phi_ret_plus_expanded}
\end{align}
for $n\in\mathcal{N}_+$, and
\begin{align}
    \phi_{s,n}(t) &\simeq \phi_{s,0} + \Omega_{s0} t - \frac{n - 3}{2}\Omega_{s0}\Delta t_0^{\rm ss} - \Omega_{s0}\left.\Delta t_0^{\rm os}\right|_{n\to n-1}
    \nonumber\\
    &+\,a\bigg[\Omega_{s1} t - \frac{n - 3}{2}\left(
    \Omega_{s0}\Delta t_{1,+}^{\rm ss} + \Omega_{s1}\Delta t_0^{\rm ss} \right) 
    \nonumber\\
    &-\,\Omega_{s0}\left.\Delta t_1^{\rm os}\right|_{n\to n-1}
    - \Omega_{s1}\left.\Delta t_0^{\rm os}\right|_{n\to n-1}
    \bigg]\,,
\label{phi_ret_minus_expanded}
\end{align}
for $n\in\mathcal{N}_-$.

\subsection{Exact retarded azimuths}
\label{sec:exact_delay}

Beyond the static-delay approximation, the propagation time of the photons forming each image must be evaluated at the retarded azimuth from which they were emitted. Let $T_n(\phi_s)$ denote the propagation time of the photons forming the image $n$ when they are emitted at source azimuth $\phi_s$. The photons forming image $n$ and the reference image $n=2$, both contributing to the visibility at the same observing time, left the source at times $t+T_2(\phi_s(t))-T_n(\phi_{s,n}(t))$ and $t$, when it was at $\phi_{s,n}(t)$ and $\phi_s(t)$, respectively. Since the source moves with angular velocity $\Omega_s$, the retarded azimuths satisfy
\begin{equation}
    \phi_{s,n}(t) = \phi_s(t) - \Omega_s\Big[T_n\big(\phi_{s,n}(t)\big) - T_2\big(\phi_s(t)\big)\Big]\,,
\label{retarded_exact}
\end{equation}
which is an implicit equation for $\phi_{s,n}(t)$. The propagation time $T_n$ can be obtained from Ref.~\cite{Aratore2024}. Its zeroth-order part is Eq.~(5.13) therein, which follows from integrating Eq.~\eqref{carter_time} from the source to the observer with $a=0$. At first order, the last line of Eq.~(6.2) therein gives the delay between two images as the difference of two single-image terms. Each of them is fixed by the spin correction to the closest-approach distance of the corresponding photon, Eq.~(6.5) therein, and provides the linear part of $T_n$. In our notation, this gives
\begin{multline}
    T_n(\phi_s) = m\left[3\sqrt{3}\,\psi_{0,\pm}(n) - K_s - K_o\right]
    \\
    +\,\,3\sqrt{3}\,a\,m\sqrt{1 - \mu_o^2}\,\tilde{\psi}_\pm(n)\cos\alpha_{0,\pm}\,,\quad n\in\mathcal{N}_\pm\,,
\label{T_n_def}
\end{multline}
where, as in Sec.~\ref{sec:spinformula}, the dependence on the image order is written explicitly. The dependence on $\phi_s$ enters only through $\alpha_{0,\pm}$, Eq.~\eqref{alpha0_pm}, which also determines $h_\pm$ and hence $\psi_{0,\pm}$ and $\tilde{\psi}_\pm$. The constants $K_s$ and $K_o$, given in Eq.~(5.14) of Ref.~\cite{Aratore2024}, depend only on $r_s$ and $r_o$, respectively, and cancel in Eq.~\eqref{retarded_exact}. At fixed $\phi_s$, the delays of Sec.~\ref{sec:strongdeflectionlimit} are differences of propagation times, namely $\Delta t^{\rm ss}_\pm=T_{n+2}-T_n$ for $n\in\mathcal{N}_\pm$ and $\Delta t^{\rm os}=T_{n+1}-T_n$ for $n\in\mathcal{N}_+$. Evaluating both propagation times in Eq.~\eqref{retarded_exact} at $\phi_s(t)$ thus recovers Eqs.~\eqref{phi_ret_plus_exact} and \eqref{phi_ret_minus}. The static-delay approximation therefore neglects the variation of $T_n$ between $\phi_s(t)$ and $\phi_{s,n}(t)$, which grows with the image order, since so does the separation between these two azimuths. For each $n$, Eq.~\eqref{retarded_exact} is a single scalar equation, which we solve numerically for $\phi_{s,n}(t)$; the result is then substituted into the visibility of Eq.~\eqref{visibility_orbiting_source}. All the visibility maps of Sec.~\ref{sec:application} are computed in this way, and the accuracy of the static-delay approximation is discussed there.

\subsection{A concrete application}
\label{sec:application}

Having derived the visibility of an orbiting compact source in general form, we now specialize to some configurations of interest. We evaluate all these configurations at $t=0$, fixing the source azimuth probed by the $n=2$ image to $\phi_{s,2}(0)=\phi_s(0)=\phi_{s,0}=\pi/4$. We recall that the source is placed on the equatorial plane, so that $\vartheta_s=\pi/2$ and $\mu_s=0$; whenever orbital motion is included, it follows the circular orbit described at the beginning of this section. For the black-hole and observer parameters, we adopt values representative of M87*, taking $m=6.5\times10^9\,M_\odot$, with $M_\odot$ denoting the solar mass, $r_o=16.5\,\mathrm{Mpc}$, and $\vartheta_o=17^\circ$. The observer is therefore effectively in the asymptotic region, with $\eta_o \simeq 1$. To complete the source model, we choose the extent of the source along each coordinate as
\begin{align}
    \Delta r_s &= \frac{r_s}{2000}\,,
\label{Deltars}
    \\
    \Delta\vartheta_s &= \Delta\phi_s
    = \frac{\Delta r_s}{r_s} = \frac{1}{2000}\,.
\label{DeltathetasDeltaphis}
\end{align}
The angular size of the source is thus the same in all the configurations considered below.

The finite angular and radial extensions of the higher-order images additionally require the derivatives entering Eqs.~\eqref{Delta_alpha0pm}, \eqref{Delta_alpha1pm}, and \eqref{Delta_epsilon_pm_def}. These derivatives propagate the characteristic source extensions in
Eqs.~\eqref{Deltars} and \eqref{DeltathetasDeltaphis} into $\Delta\alpha_\pm$ and
$\Delta\theta_\pm$ through Eqs.~\eqref{Delta_alpha_pm_expansion} and \eqref{Delta_theta_pm_def}, respectively. Their explicit expressions for an equatorial source, evaluated at $\mu_s=0$, are collected in Appendix~\ref{app:derivatives}. 

We now compare four configurations, distinguished by the presence or absence of black-hole spin and source orbital motion:
\begin{enumerate}
    \item a static source in Schwarzschild spacetime;
    \item an orbiting source in Schwarzschild spacetime;
    \item a static source in Kerr spacetime;
    \item an orbiting source in Kerr spacetime.
\end{enumerate}
Accordingly, we postpone assigning numerical values to the spin parameter $a$ and the source radius $r_s$. The former distinguishes the Schwarzschild and Kerr configurations, while the latter controls both the finite-distance lensing geometry and the relevance of the orbital motion. Indeed, the difference between the reference and retarded
source azimuths is of order $\Omega_s\Delta t$, where $\Delta t$ is the propagation-time difference associated with a given higher-order image. Since $\Omega_{s0}\propto r_s^{-3/2}$, this displacement becomes increasingly small as the source radius grows. Consequently, for sufficiently large $r_s$, different image orders probe nearly the same source azimuth and the effects of orbital motion become negligible. More specifically, the four configurations we consider correspond respectively to
\begin{equation*}
    \begin{array}{r@{\hspace{1.5em}}l}
        1. & a = 0\,, \quad \Omega_s=0\,; \\[0.3em]
        2. & a = 0\,, \quad \Omega_s \neq 0\,; \\[0.3em]
        3. & a \neq 0\,, \quad \Omega_s = 0\,; \\[0.3em]
        4. & a \neq 0\,, \quad \Omega_s \neq 0\,.
\end{array}
\end{equation*}
For each configuration, we retain the first three opposite-side image pairs, truncating the two half-orbit sequences in Eq.~\eqref{Npm_def} to
\begin{align}
    \mathcal N_+^{(3)} &= \{2,4,6\}\,,
    \\
    \mathcal N_-^{(3)} &= \{3,5,7\}\,.
\end{align}
The retained pairs are therefore $(2,3)$, $(4,5)$, and $(6,7)$.

Before presenting the resulting visibility maps, a brief clarification may be useful here.  In the spherically symmetric case considered in Ref.~\cite{Aratore2021}, all images of a static source lie on a common line on the observer screen. One may therefore choose the screen axes along the radial and tangential image directions, with the conjugate spatial-frequency components $u$ and $v$ probing these directions, respectively. At the level of approximation considered in Ref.~\cite{Aratore2021}, the tangential width is common to all higher-order images, so the $v$-dependent Gaussian factor can be extracted from the image sum. The nontrivial hierarchy of higher-order images can consequently be represented by a one-dimensional visibility profile as a function of $u$.

Away from the static Schwarzschild limit, this simplification is lost for two independent reasons. In Kerr spacetime, the spin corrections make the image position angles dependent on the image order \cite{Aratore2024}. Moreover, when the source is in orbital motion, different image orders probe different retarded source azimuths $\phi_{s,n}(0)$. Thus, either spin or orbital motion --- and, in general, their combined action --- causes the local radial and tangential directions to vary from one image to another. Consequently, the functions $\mathcal{R}_\pm(u,v)$ and $\mathcal{J}_\pm(u,v)$ entering the visibility depend nontrivially on both spatial-frequency components, and the visibility must be represented over the full two-dimensional $(u,v)$-plane.

As already mentioned in Secs.~\ref{sec:intro} and \ref{sec:visibilityfunction}, successive higher-order images are progressively thinner and therefore
leave their interferometric signatures at increasingly large spatial frequencies. Thus, to display this hierarchy within a single two-dimensional map, we adopt a logarithmic parametrization of the baseline plane. To avoid cluttering the notation, we continue to denote the baseline coordinates by $(u,v)$. The visibility is evaluated after the replacement
\begin{equation}
    (u,v) \longrightarrow \frac{r_o}{2m}\,10^{\sqrt{u^2 + v^2}}\left(\frac{u}{\sqrt{u^2 + v^2}},\frac{v}{\sqrt{u^2 + v^2}}\right)\,,
\label{logarithmic_baseline_map}
\end{equation}
where $(u,v)\neq(0,0)$. The factor $r_o/(2m)$, namely the inverse of the angular scale $2m/r_o$, sets the unit of the physical spatial frequencies. Since, for fixed $a/m$ and $r_s/m$, $\theta_\pm$, $\Delta\theta_\pm$, and $\delta$ are proportional to $m/r_o$, while $\alpha_\pm$ and $\Delta\alpha_\pm$ do not depend on $m$ and $r_o$, the normalized maps introduced below are independent of the black-hole mass and distance. The logarithm compresses the wide range of baselines required to resolve successive higher-order images, whose thicknesses decrease exponentially with $n$, while preserving the orientation of each baseline.

Moreover, to compare the visibility patterns independently of the total flux contributed by the higher-order images retained in the sum, we normalize the visibility amplitude by its zero-baseline value, and define
\begin{equation}
    \bar{V}(u,v) \coloneqq \left|\frac{V(u,v;0)}{V(0,0;0)}\right|\,.
\label{normalizedvisibility}
\end{equation}
Indeed, at zero baseline, the Fourier phase factor appearing in Eq.~\eqref{visibility_def_polar} is equal to unity, so that $V(0,0;0)$ reduces to the integral of the image brightness over the observer's sky and therefore gives the total flux of the modeled higher-order images at $t=0$. The normalization consequently sets $\bar{V}(0,0)=1$ and expresses the visibility amplitude relative to this total flux.

\begin{figure*}[p]
    \centering
    \subfloat[Schwarzschild, fixed source.
    \label{fig:Schwarzschild}]{\includegraphics[width=0.42\textwidth]{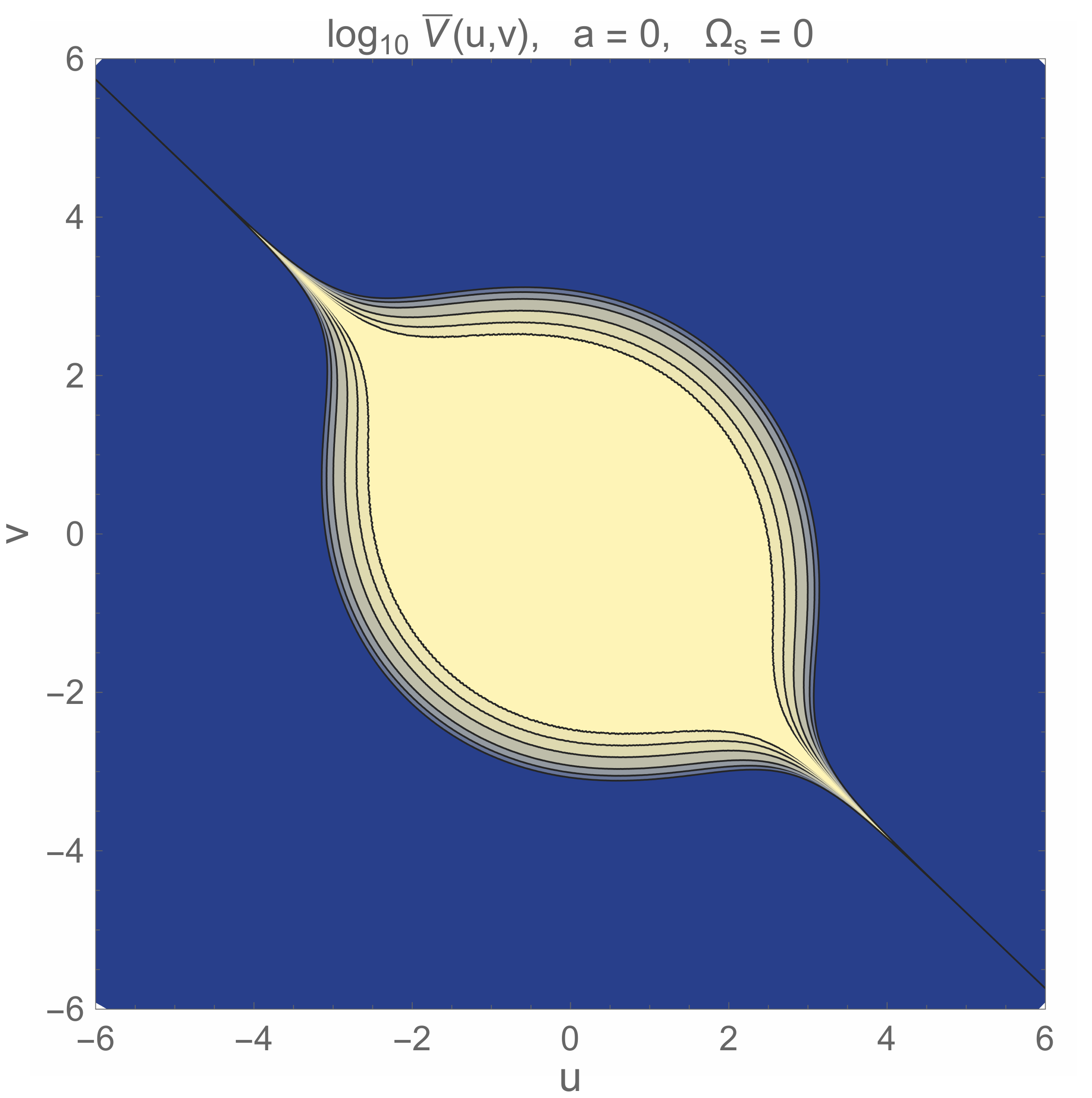}}
    \hspace{0.01\textwidth}
    \subfloat[Schwarzschild, source in orbital motion. \label{fig:Schwarzschild_orbitalmotion}]{\includegraphics[width=0.42\textwidth]{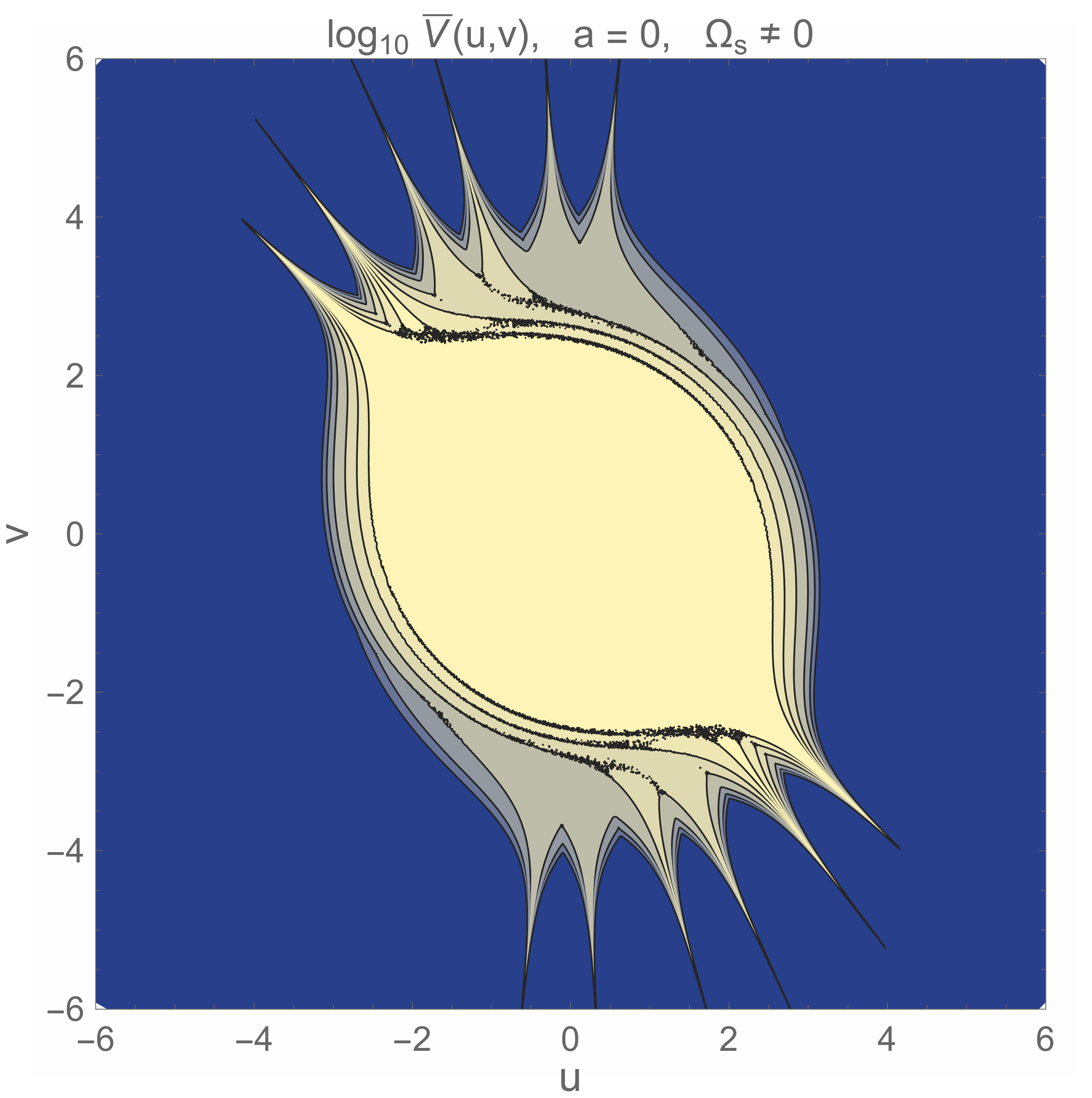}}

    \vspace{0.4cm}

    \subfloat[Kerr, fixed source.
    \label{fig:Kerr}]{\includegraphics[width=0.42\textwidth]{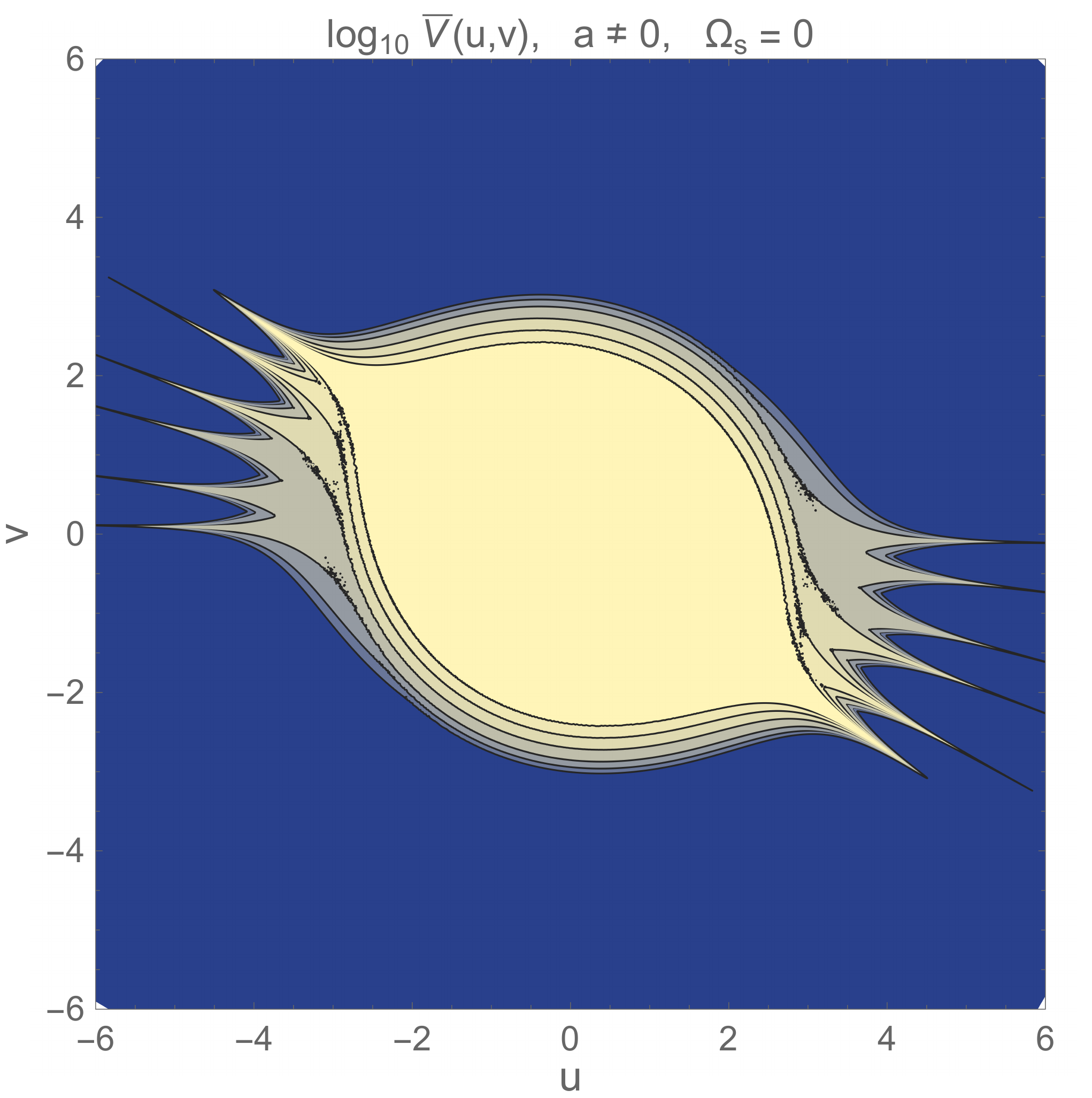}}
    \hspace{0.01\textwidth}
    \subfloat[Kerr, source in orbital motion.
    \label{fig:Kerr_orbitalmotion}]{\includegraphics[width=0.42\textwidth]{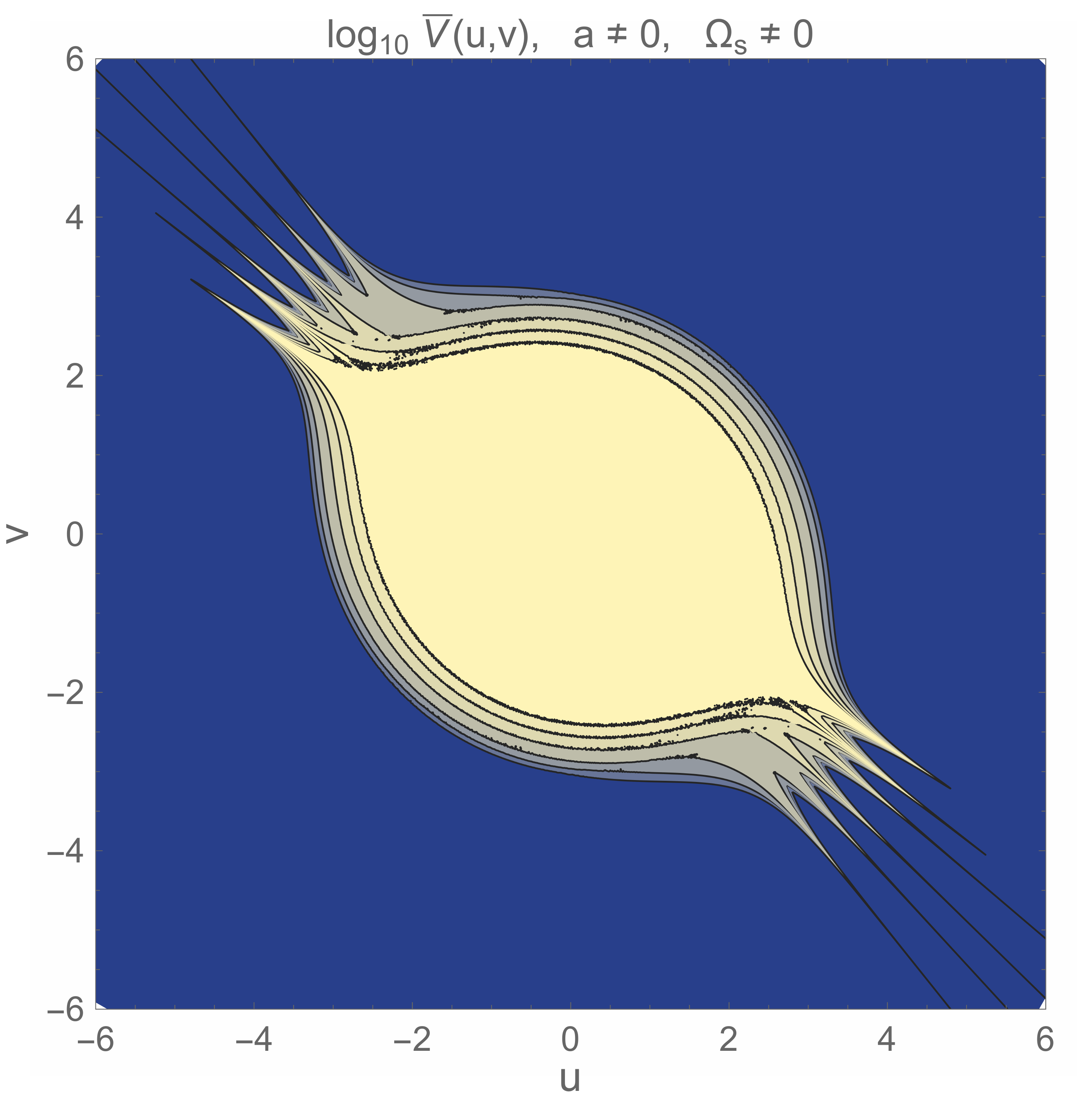}} \caption{Maps of the normalized visibility amplitude for the four configurations considered in the text. All panels are evaluated at $t=0$ for a source with $\vartheta_s=\pi/2$, $\mu_s=0$, and azimuth probed by the $n=2$ image $\phi_{s,2}(0)=\phi_{s,0}=\pi/4$. The black-hole and observer parameters are $m=6.5\times10^9M_\odot$, $r_o=16.5\,\mathrm{Mpc}$, and $\vartheta_o=17^\circ$, so that $\eta_o\simeq1$. In the static configurations, panels (a) and (c), $r_s=2000m$ and $\Omega_s=0$; in the orbiting configurations, panels (b) and (d), $r_s=20m$ and the source follows the prograde circular orbit of Sec.~\ref{sec:intsignatureimagesorbitingemitter}. The source has $\Delta\vartheta_s=\Delta\phi_s=\Delta r_s/r_s=1/2000$ in every panel, i.e., $\Delta r_s=m$ in panels (a) and (c) and $\Delta r_s=m/100$ in panels (b) and (d). For the Kerr configurations, panels (c) and (d), $a/m=0.1$. In the orbiting cases, each image contribution is evaluated at its order-dependent retarded source azimuth $\phi_{s,n}(0)$, obtained from Eq.~\eqref{retarded_exact}. The visibility sum includes the first three opposite-side pairs, $(2,3)$, $(4,5)$, and $(6,7)$. The plotted quantity is $\log_{10}\bar V(u,v)$, with $(u,v)$ the rescaled baseline coordinates of Eq.~\eqref{logarithmic_baseline_map}. The black contour lines mark the levels $\log_{10}\bar V=-1,-2,-4,-8,-12$, and $-16$, corresponding respectively to $\bar V=10^{-1},10^{-2},10^{-4},10^{-8},10^{-12}$, and $10^{-16}$. Spikes narrower than the contour lines are continued as lines up to the point where $\bar V$ drops to $10^{-16}$. For the higher-order images with $n\geq5$ in panel (a) and $n\geq4$ in panels (b)--(d), this point lies outside the plotted region, and the corresponding spikes are cut at its boundary. The decimal logarithm displays the visibility-amplitude range, while the independent radial rescaling of Eq.~\eqref{logarithmic_baseline_map} compresses the broad range of physical baseline lengths and preserves their orientation. Note that Eq.~\eqref{logarithmic_baseline_map} is undefined at the origin of the rescaled $(u,v)$-plane. Any color rendered at the single point $(u,v)=(0,0)$ therefore has no physical interpretation and does not represent zero baseline; the physical value $|V(0,0;0)|$ enters only as the normalization in Eq.~\eqref{normalizedvisibility}.}
    \label{fig:visibility_maps}
\end{figure*}

Figure~\ref{fig:visibility_maps} shows the visibility maps corresponding to configurations 1--4; each panel displays $\log_{10}\bar V(u,v)$. We set $r_s=2000m$ for the static sources, $r_s=20m$ for the orbiting sources, and $a/m=0.1$ in the Kerr cases.

The four maps in Fig.~\ref{fig:visibility_maps} share a common morphology, consisting of a nearly rounded central region surrounded by narrow spikes extending toward large rescaled baseline radii. For the sources considered here, the condition $\pi\theta_\pm B\,\Delta\alpha_\pm^2 \ll 1$ discussed after
Eq.~\eqref{visibility_final} holds over most of the plotted range; only the outermost parts of the spikes, near the edges of the panels, are expected to be dimmed and broadened by the curvature of the arcs.

Although positive- and negative-parity images form on opposite sides of the black hole on the observer's sky, this distinction does not correspond to the two sides of the visibility-amplitude map. Since the brightness distribution is real, the visibility satisfies $\bar V(-u,-v)=\bar V(u,v)$, and each individual image therefore produces a pair of antipodal spikes. The contributions of both parity sequences are consequently present on both sides of each visibility map.

We begin with the static Schwarzschild configuration shown in Fig.~\ref{fig:Schwarzschild}. As discussed in Ref.~\cite{Aratore2021}, spherical symmetry places all the higher-order images of a static source along a common direction on the observer screen. Their visibility signatures are consequently aligned along a common axis in the $(u,v)$-plane, so that the six retained images produce only one pair of superposed antipodal spikes. A one-dimensional cut along this axis recovers the staircase-like
visibility profile discussed in Ref.~\cite{Aratore2021}.

Figure~\ref{fig:Schwarzschild_orbitalmotion} shows the effect of source orbital motion in Schwarzschild spacetime. In this case, the different image orders no longer correspond to the same source position because the photons forming them have different propagation times. According to Eq.~\eqref{retarded_exact}, each image is therefore determined by the source position at its order-dependent retarded azimuth $\phi_{s,n}(0)$. For the prograde orbit considered here, photons with larger half-orbit number were emitted at progressively earlier orbital phases and consequently probe progressively smaller source azimuths.

For the equatorial source considered here,
Eq.~\eqref{eq:dalpha0-dphis} shows that
$\partial\alpha_{0,\pm}/\partial\phi_s>0$, since $\mu_o=\cos\vartheta_o>0$. The progressively smaller retarded source azimuths therefore correspond to progressively smaller image position angles. Since $\alpha$ increases counterclockwise on the observer screen, the spikes rotate clockwise with increasing image order, as seen in Fig.~\ref{fig:Schwarzschild_orbitalmotion}.

We also note that the angular spacing between consecutive spikes is not uniform. This behavior can be understood from the retarded-azimuth prescriptions in Eqs.~\eqref{phi_ret_plus_exact} and \eqref{phi_ret_minus_from_plus} of the static-delay approximation, which is accurate here for two reasons. First, without the spin term of Eq.~\eqref{T_n_def}, whose weight grows with $n$, $T_n$ depends on the source azimuth only through $h_\pm$, which is the same for all the images of a given parity. Second, the separations between consecutive spikes, produced by the orbital motion alone, are large compared with the errors introduced by the approximation. Consider an opposite-side pair $(n,n+1)$, with $n$ even. The positive- and negative-parity images in this pair are separated by the opposite-side propagation delay $\Delta t^{\rm os}$ defined in Eq.~\eqref{dt_opp_n}. The next positive-parity image, labeled by $n+2$, is instead separated from the image $n$ by the same-side delay $\Delta t^{\rm ss}_+$ defined in Eq.~\eqref{dt_same_n}. Consequently, the propagation-time separation between the negative-parity image $n+1$ and the following positive-parity image $n+2$ is given by the difference between the same-side and opposite-side delays. For the configuration shown in Fig.~\ref{fig:Schwarzschild_orbitalmotion}, the propagation-time separation within each opposite-side pair is smaller than that between the negative-parity image and the following positive-parity image. Since the source moves uniformly along its circular orbit, the smaller propagation-time separation corresponds to a smaller separation between the retarded source azimuths, whereas the larger one corresponds to a larger azimuthal separation. After these retarded source positions are mapped into image position angles through Eq.~\eqref{alpha0_pm}, this alternating sequence produces the narrower and wider angular gaps visible between successive spikes. The precise sizes of these gaps are also affected by the nonlinear dependence of $\alpha_{0,\pm}$ on the source azimuth $\phi_s$.

We proceed by considering the static Kerr configuration in Fig.~\ref{fig:Kerr}. Since the source is fixed, all image orders are evaluated at the same source azimuth, and the separation of the spikes is entirely due to the spin-dependent corrections to the image position angles. Black-hole rotation modifies the propagation of near-critical photons through frame dragging, and the accumulated correction increases with the number of half-orbits performed around the black hole. Consequently, successive images belonging to the same parity sequence no longer share the same position angle. Their difference is given, to first order in the spin parameter, by Eq.~\eqref{diffalpha_same_side}. For $a>0$ and $\mu_o>0$, this quantity is positive, so the image position angles increase with $n$. The corresponding spikes therefore progress counterclockwise with increasing image order. Their ordering is opposite to the clockwise displacement generated by the retarded source motion in the Schwarzschild case.

Finally, Fig.~\ref{fig:Kerr_orbitalmotion} includes both effects. The lensing quantities associated with each image are evaluated at the corresponding retarded source azimuth $\phi_{s,n}(0)$, while the image positions also contain their linear-in-spin corrections. Source motion tends to decrease the image position angles and rotate the spike sequence clockwise, whereas the Kerr correction tends to increase them and rotate the sequence counterclockwise. The morphology of Fig.~\ref{fig:Kerr_orbitalmotion} is therefore determined by the competition between these two effects.

For the chosen parameters, the displacement produced by source motion is larger than the frame-dragging contribution; the overall progression of the spikes is therefore clockwise, although it is partly opposed and distorted by the spin correction. This behavior is not universal. Its magnitude and direction depend on the black-hole spin, the source radius and angular velocity, the sense of the orbit, and the observer inclination. Different choices of these parameters may, e.g., weaken the relative displacement or produce an approximate cancellation. The static-delay approximation introduced in Sec.~\ref{sec:static_delay} captures the overall morphology of Fig.~\ref{fig:Kerr_orbitalmotion}, namely the ordering, lengths, and clockwise progression of the spikes. However, the competition between orbital motion and frame dragging brings consecutive spikes in the visibility map closer together than either effect alone would, so that even the small errors of this approximation alter the relative widths of the gaps between them, whose correct description therefore requires the exact solution.

The radial length of the spikes has a simple physical interpretation. Each image is a thin arc, whose length $\theta_\pm|\Delta\alpha_\pm|$ is much larger than its thickness $\Delta\theta_\pm$. Since a baseline of length $B$ (in wavelengths) resolves angular scales $\sim1/B$ along its direction, the image is resolved at $B\sim1/(\theta_\pm|\Delta\alpha_\pm|)$ for most baseline orientations, whereas along $\alpha_\pm$, perpendicular to the arc, it requires the much longer $B\sim1/\Delta\theta_\pm$, which is where the spike ends (for the spikes reaching the edges of the panels in Fig.~\ref{fig:visibility_maps}, this occurs beyond the plotted range). The factor $e^{-\psi_{0,\pm}}$ produces an approximate $e^{2\pi}$ thinning per additional loop, extending the spikes to longer baselines, with corrections from spin and, for an orbiting source, from differing retarded positions. Their growth with $n$ thus reflects the exponential demagnification of successive higher-order images.

Because the spike lengths are set by $\Delta\theta_\pm$, they depend not only on the angular size of the source, which is the same in all panels, but also on its radius, which differs between the static and orbiting configurations. The angular ordering of the spikes thus compares the effects of spin and orbital motion, whereas their lengths do not.

\section{Summary and conclusions}
\label{sec:conclusions}

In this work, we have developed an analytical framework for computing the interferometric signatures of higher-order images produced by a compact source in a slowly rotating Kerr spacetime. Starting from the strong-deflection-limit expressions for the angular positions and propagation times of the images reviewed in Sec.~\ref{sec:strongdeflectionlimit}, we extended the point-source description to a source of finite size. By evaluating the first-order variations of the image positions with respect to the source coordinates, we determined the radial thickness and angular extension of each higher-order image and modeled its local brightness profile as a Gaussian distribution in polar coordinates centered on the shadow. This construction led, in Sec.~\ref{sec:visibilityfunction}, to the analytical expression for the complex visibility given in Eq.~\eqref{visibility_final}, valid at linear order in the black-hole spin. The functions $\mathcal{R}_\pm$ and $\mathcal{J}_\pm$ entering this expression control, respectively, the decay of the visibility amplitude and its interferometric phase, while retaining the dependence on the position and finite size of each image.

For a static source in the equatorial plane, we derived in Sec.~\ref{sec:spinformula} a direct relation between the dimensionless Kerr spin parameter and the angular displacement of two successive images belonging to the same parity sequence. Equation~\eqref{spinformula} expresses the spin in terms of the difference between the position angles of the images with half-orbit numbers $n$ and $n+2$, providing a direct geometrical relation between the black-hole spin and the relative angular positions of higher-order images. This result applies to a static source, for which all images within a given parity sequence correspond to the same source position, and does not directly extend to a moving emitter.

In Sec.~\ref{sec:intsignatureimagesorbitingemitter}, we extended the construction to a compact source moving on a circular equatorial orbit. For an orbiting source, the higher-order images observed at a given time are formed by photons emitted at different orbital phases. We accounted for this by evaluating the lensing quantities associated with each image at the corresponding retarded source azimuth. We derived closed-form expressions for these azimuths in the static-delay approximation, while the visibility maps were computed with the exact azimuths, obtained by solving a single scalar equation numerically for each image. The resulting time-dependent visibility is given by Eq.~\eqref{visibility_orbiting_source}, which incorporates the orbital motion of the source while keeping all the lensing quantities in closed form.

We illustrated the resulting interferometric signatures by comparing four configurations:~a static and an orbiting source in Schwarzschild spacetime, and a static and an orbiting source in Kerr spacetime. In the static Schwarzschild case, spherical symmetry implies that all higher-order images are aligned along a single line on the observer screen. Their visibility contributions therefore form a single pair of superposed antipodal spikes. The corresponding one-dimensional visibility profile along the common axis of the spikes reproduces the staircase-like structure obtained in the spherically symmetric analysis of Ref.~\cite{Aratore2021}.

The orbital motion of the source breaks this alignment even in Schwarzschild spacetime, since the higher-order images correspond to different retarded source positions. For the prograde configuration
and screen convention adopted here, the associated spikes progress clockwise with increasing image order. Their angular separations are not uniform, reflecting the alternating propagation-time intervals between the two members of an opposite-side pair and between consecutive pairs.

For a static source in Kerr spacetime, the angular separation of the spikes is instead produced entirely by the spin corrections to the image positions. In the configuration considered here, frame dragging causes the spikes to progress counterclockwise with increasing image
order, in the direction opposite to that produced by source motion. When both effects are present, the visibility pattern is therefore determined by their competition. For the parameters adopted in Sec.~\ref{sec:application}, the displacement generated by source motion is larger, so that the overall progression remains clockwise but is partially opposed and distorted by the Kerr correction. This behavior is not universal, since the relative magnitude and direction of the two effects depend on the black-hole spin, the source radius and angular velocity, the sense of the orbit, and the observer inclination. The angular ordering of the long-baseline features provides the most
direct comparison between frame dragging and source motion.

Some extensions of the present analysis can be considered. A first direction would be to apply the present framework to parametrized stationary and axisymmetric black-hole spacetimes, such as the Konoplya--Rezzolla--Zhidenko parametrization \cite{KonoplyaRezzollaZhidenko2016}. This would make it possible to determine how the deviation parameters modify the positions and propagation times of higher-order images and, consequently, the amplitude and phase of the complex visibility. Such an analysis would generalize the results of Ref.~\cite{Feleppa2025} beyond spherical symmetry.

A second direction would be to extend the analysis beyond the perturbative low-spin regime considered here. A strong-deflection-limit analysis applicable to extremal rotating black holes was recently developed in Ref.~\cite{Feleppaextremal2026}. Generalizing this construction to sources at finite distance and subsequently relaxing the extremal hypothesis to near-extremal black holes would in principle provide the lensing quantities required to derive the complex visibility in the high-spin regime. This regime is particularly relevant astrophysically: many accreting supermassive black holes are found to be rapidly spinning \cite{Reynolds2021}, and a high spin has also been estimated for M87* \cite{Tamburini1}.

\section*{Acknowledgments} \label{sec:acknowledgements}

FF gratefully acknowledges the hospitality of the Center of Gravity at the Niels Bohr Institute, where part of this work was carried out. FF acknowledges support from the Istituto Nazionale di Fisica Nucleare (INFN) through the Commissione Scientifica Nazionale 4 (CSN4) Iniziativa Specifica ``Quantum Universe'' (QGSKY).

\onecolumngrid
\appendix
\section{Relevant derivatives}
\label{app:derivatives}

In this appendix, we collect the derivatives of the lensing quantities $\alpha_{0,\pm}$, $\alpha_{1,\pm}$, $\epsilon_{0,\pm}$, and $\epsilon_{1,\pm}$, introduced in Sec.~\ref{sec:strongdeflectionlimit}, with respect to the source variables $\eta_s$, $\mu_s$, and $\phi_s$. Through Eqs.~\eqref{Delta_alpha_pm_def2} and \eqref{Delta_epsilon_pm_def}, these derivatives determine the angular and radial extensions of the higher-order images produced by a source of finite size, and hence the functions $\mathcal{R}_\pm$ and $\mathcal{J}_\pm$ entering the complex visibility of Eq.~\eqref{visibility_final}. For the circular equatorial orbit considered in this work, $\vartheta_s=\pi/2$, and all derivatives reported below are therefore evaluated at $\mu_s=0$. Where a double sign appears, the upper (lower) sign refers to the positive-parity (negative-parity) branch. We list first the derivatives entering $\Delta\alpha_\pm$ and then those entering $\Delta\epsilon_{i,\pm}$, in each case at zeroth and at first order in the spin. Since $\alpha_{0,\pm}$ does not depend on $\eta_s$, the nonvanishing derivatives entering $\Delta\alpha_{0,\pm}$ are
\begin{align}
    \frac{\partial \alpha_{0,\pm}}{\partial \mu_s} &= -\frac{\sqrt{1 - \mu_o^2}\,\sin\phi_s}{1 - (1 - \mu_o^2)\cos^2\phi_s}\,,
\label{eq:dalpha0-dmus}
    \\[1ex]
    \frac{\partial \alpha_{0,\pm}}{\partial \phi_s} &= \frac{\mu_o}
    {1 - (1 - \mu_o^2)\cos^2\phi_s}\,.
\label{eq:dalpha0-dphis}
\end{align}
The derivatives entering the first-order contribution $\Delta\alpha_{1,\pm}$ are
\begin{align}
    \frac{\partial \alpha_{1,\pm}}{\partial \eta_s} &= -\frac{2\mu_o}
    {m\sqrt{3 - 2\eta_s}\,(1 + 2\eta_s)\left[1 - (1 - \mu_o^2)\cos^2\phi_s\right]}\,,
\label{eq:dalpha1-detas}
    \\[1ex]
    \frac{\partial \alpha_{1,\pm}}{\partial \mu_s} &= \frac{(1-\mu_o^2)\sin^2\phi_s - \mu_o^2}{\left[1 - (1 - \mu_o^2)\cos^2\phi_s\right]^{3/2}}\left[\pm\frac{2}{3\sqrt{3}\,m} + \frac{\sqrt{1 - \mu_o^2}\,\cos\phi_s}
    {\sqrt{1 - (1 - \mu_o^2)\cos^2\phi_s}}\,
    \tilde{\psi}_{\pm}\right]\,,
\label{eq:dalpha1-dmus}
    \\[1ex]
    \frac{\partial \alpha_{1,\pm}}{\partial \phi_s} &= -\frac{2\mu_o\sqrt{1-\mu_o^2}\,\sin\phi_s}{\left[1 - (1 - \mu_o^2)\cos^2\phi_s\right]^{3/2}}\left[\pm\frac{2}{3\sqrt{3}\,m} + \frac{\sqrt{1 - \mu_o^2}\,\cos\phi_s}
    {\sqrt{1 - (1 - \mu_o^2)\cos^2\phi_s}}
    \tilde{\psi}_{\pm}\right]\,.
\label{eq:dalpha1-dphis}
\end{align}
The derivatives entering $\Delta\epsilon_{0,\pm}$ are
\begin{align}
    \frac{\partial \epsilon_{0,\pm}}{\partial \eta_s} &= \epsilon_{0,\pm}\frac{\sqrt{3}}
    {\eta_s\sqrt{3 - 2\eta_s}}\,,
\label{eq:depsilon0-detas}
    \\[1ex]
    \frac{\partial \epsilon_{0,\pm}}{\partial \mu_s} &= \pm\epsilon_{0,\pm}\frac{\mu_o}
    {\sqrt{1 - (1 - \mu_o^2)\cos^2\phi_s}}\,,
    \label{eq:depsilon0-dmus}
    \\[1ex]
    \frac{\partial \epsilon_{0,\pm}}{\partial \phi_s} &= \pm\epsilon_{0,\pm}\frac{\sqrt{1 - \mu_o^2}\,\sin\phi_s}
    {\sqrt{1 - (1 - \mu_o^2)\cos^2\phi_s}}\,.
\label{eq:depsilon0-dphis}
\end{align}
Finally, the derivatives entering the first-order contribution $\Delta\epsilon_{1,\pm}$ are
\begin{align}
    \frac{\partial \epsilon_{1,\pm}}{\partial \eta_s} &= \pm\epsilon_{0,\pm}\frac{\sqrt{1-\mu_o^2}\,\sin\phi_s}{\sqrt{1 - (1 - \mu_o^2)\cos^2\phi_s}}\left[
    \frac{\sqrt{3}\,\tilde{\psi}_{\pm}}{\eta_s\sqrt{3-2\eta_s}} - \frac{2}
    {m\sqrt{3-2\eta_s}\,(1 + 2\eta_s)}\right]\,,
\label{eq:depsilon1-detas}
    \\[1ex]
    \frac{\partial \epsilon_{1,\pm}}{\partial \mu_s} &= \epsilon_{0,\pm}\,\mu_o\sqrt{1 - \mu_o^2}\,\sin\phi_s\left\{\frac{1}
    {1 - (1 - \mu_o^2)\cos^2\phi_s}\left(\tilde{\psi}_{\pm}-\frac{2}{3\sqrt{3}\,m}\right) \mp \frac{\sqrt{1-\mu_o^2}\,\cos\phi_s}{\left[1 - (1 - \mu_o^2)\cos^2\phi_s\right]^{3/2}}\,\tilde{\psi}_{\pm}
    \right\}\,,
\label{eq:depsilon1-dmus}
    \\[1ex]
    \frac{\partial \epsilon_{1,\pm}}{\partial \phi_s} &= \epsilon_{0,\pm}\sqrt{1 - \mu_o^2}\left\{\frac{\sqrt{1 - \mu_o^2}\,\sin^2\phi_s}{1 - (1 - \mu_o^2)\cos^2\phi_s}
    \left(\tilde{\psi}_{\pm}-\frac{2}{3\sqrt{3}\,m}\right) \pm \frac{\mu_o^2\cos\phi_s}
    {\left[1 - (1 - \mu_o^2)\cos^2\phi_s\right]^{3/2}}\,\tilde{\psi}_{\pm}\right\}\,.
\label{eq:depsilon1-dphis}
\end{align}

\twocolumngrid

\end{document}